%% file: quantum_version.tex
\documentclass[a4paper,aps,prx,10pt,twocolumn,superscriptaddress,accepted=2021-07-10]{quantumarticle}
\pdfoutput=1
\usepackage[ascii]{inputenc}
\usepackage{amsmath,amssymb,amsfonts,amsthm}
\usepackage{tikz}
\usepackage{graphicx}
\usepackage{braket}
\usepackage[caption=false]{subfig}
\usepackage[pdftex,bookmarks=false,colorlinks=true,linkcolor=blue,
citecolor=blue,filecolor=black,urlcolor=blue]{hyperref}

\newcommand{\ud}{\mathrm{d}}

\newcommand{\R}{\mathbb{R}}

\DeclareMathOperator{\e}{e}

\newcommand{\imag}{\ensuremath{\mathrm{i}}}

\DeclareMathOperator*{\argmin}{arg\,min}

\begin{document}

\title{Quantum Algorithms for Solving Ordinary Differential Equations via Classical Integration Methods}

\author{Benjamin Zanger} \orcid{0000-0001-8511-8160}
\email{benjamin.zanger@tum.de}
\affiliation{Technical University of Munich, Department of Informatics, Boltzmannstra{\ss}e 3, 85748 Garching, Germany}

\author{Christian B.~Mendl}\orcid{0000-0002-6386-0230}
\email{christian.mendl@tum.de}
\affiliation{Technical University of Munich, Department of Informatics, Boltzmannstra{\ss}e 3, 85748 Garching, Germany}
\affiliation{TUM Institute for Advanced Study, Lichtenbergstra{\ss}e 2a, 85748 Garching, Germany}

\author{Martin Schulz}
\email{schulzm@in.tum.de}
\affiliation{Technical University of Munich, Department of Informatics, Boltzmannstra{\ss}e 3, 85748 Garching, Germany}
\affiliation{Leibniz Supercomputing Centre, Boltzmannstra{\ss}e 1, 85748 Garching, Germany}

\author{Martin Schreiber}\orcid{0000-0002-2390-6716}
\email{martin.schreiber@tum.de}
\affiliation{Technical University of Munich, Department of Informatics, Boltzmannstra{\ss}e 3, 85748 Garching, Germany}


\begin{abstract}
Identifying computational tasks suitable for (future) quantum computers is an active field of research.
Here we explore utilizing quantum computers for the purpose of solving differential equations.
We consider two approaches: (i) basis encoding and fixed-point arithmetic on a digital quantum computer, and (ii) representing and solving high-order Runge-Kutta methods as optimization problems on quantum annealers.
As realizations applied to two-dimensional linear ordinary differential equations, we devise and simulate corresponding digital quantum circuits.
We also implement and run a 6\textsuperscript{th} order Gauss-Legendre collocation method on a D-Wave 2000Q system, showing good agreement with the reference solution.
We find that the quantum annealing approach exhibits the largest potential for high-order implicit integration methods.
As promising future scenario, the digital arithmetic method could be employed as an ``oracle'' within quantum search algorithms for inverse problems.
\end{abstract}

\maketitle

\section{Introduction}

Solving differential equations (DEs) is a ubiquitous task in the scientific and engineering community.
Traditional numerical algorithms need to integrate the time steps sequentially due to their interdependence, which is at odds with the trend towards parallelization in modern highly-parallel high-performance computing (HPC) architectures.
This trend already motivated research in various directions, such as parallel-in-time algorithms~\cite{Gander2015_Review}.
An alternative route could emerge from quantum computers, where we could benefit from better asymptotic scaling properties compared to classical computers when applied to certain tasks~\cite{shor1997polynomial, grover1997quantum, simon1997power, Arute2019}.
Here we explore approaches of utilizing quantum computers for the purpose of solving DEs: (i) using basis encoding to describe the state of the differential equation, and mimicking arithmetic operations of classical algorithms on a digital quantum computer (Section~\ref{sec:dqc_time_integration}); and (ii) reformulating the time integration as a quantum annealing problem (Section~\ref{sec:qa_time_integration}).

To realize (i), we will first design quantum circuits for arithmetic operations in numerical fixed-point representation, and then apply these to solve linear ordinary differential equations (ODEs) in two dimensions.

Our second approach (ii) reformulates Runge-Kutta integration methods (both explicit and implicit) as minimization problems suitable for a quantum annealing framework. We additionally design a flexible number representation to reach high accuracy. We then run the annealing task (based on a method of order six) on a D-Wave 2000Q system, and are able to demonstrate a good agreement with the reference solution obtained on a classical computer.

Related work includes proposals for quantum algorithms that are capable of solving both linear partial differential equations (PDEs)~\cite{HHL, Xin2018, Costa2019} and nonlinear PDEs~\cite{lubasch2020variational, lloyd2020quantum, liu2020efficient}.
While for linear differential equations an exponential advantage in the resources is known for some time, exponential advantages for nonlinear differential equations has only been found recently.
Both algorithms, for the linear and nonlinear case, assume that the state of the PDE is amplitude encoded (i.e., via the amplitudes of the quantum wavefunction), which allows a logarithmic scaling of quantum resources for an increase in the dimension of the state space.
However, this also requires state preparation for the input and quantum state tomography for measuring the output, which is imposing scaling issues.
There have been efforts to improve these scaling for certain PDEs, e.g. for the wave equation in~\cite{Costa2019}.
Liu et al.~\cite{liu2020efficient} approached this problem by embedding the initial state of nonlinear PDEs into a higher dimensional state and to derive a measurement success probability of their algorithm, enabling a better scaling for the readout with the use of amplitude amplification.
Unfortunately, their algorithm only works for dissipative differential equations.
To our knowledge, there currently does not exist any algorithm which solves the state preparation and readout problem for general PDEs.

In contrast, our work does not provide a logarithmic scaling of quantum resources with the state space but does not obey the state preparation and readout problem.
For our second approach (ii), the time complexity for a single time step in our algorithm does not scale with the state space.
Potential applications include all kind of linear and nonlinear ODEs which are expressible with low order polynomial terms, including problems from biology and fluid dynamics.

\section{Classical Time Integration Methods}
\label{sec:classical_numerical_time_integration}

This section recapitulates the numerical time integration schemes that are used in the quantum algorithms discussed later.
Specifically, we will discuss the Runge-Kutta methods and how they can be phrased as optimization problems.
Since we can only cover a few aspects of time integration, we solely target single-step methods and refer interested readers to Durran~\cite{Durran2010}, Hairer et al.~\cite{Hairera, Hairer1991} and Shu and Osher~\cite{Shu}.

Concretely, we consider a system of ordinary differential equations written as
\begin{equation}
\label{eq:ode_starter}
\frac{\ud}{\ud t} u(t) = f(u(t), t)
\end{equation}
with $u(t) \in \R^N$ for all $t \ge 0$ and the initial condition $u(0) = u_0$.

The Runge-Kutta (RK) formulation unifies various kinds of explicit and implicit time integration schemes.
Each particular RK method can be described by a Butcher table, consisting of a matrix $A \in \mathbb{R}^{s \times s}$ and two vectors, $b, c \in \mathbb{R}^s$, where $s$ is called the number of stages of the method \cite{butcher1963}.
The RK formulation for a time step $\Delta t$ is then given by the equations ($i \in \{ 1, \dots, s \}$)
\begin{equation}
\label{eq:rk_stage_equation}
    k_i = f\bigg(\tilde{u}(t) + \Delta t \sum_{j=1}^{s} A_{i j} k_j, t + \Delta t c_i\bigg)
\end{equation}
and
\begin{equation}
\label{eq:rk_closure_equation}
    \tilde{u}(t + \Delta t) = \tilde{u}(t) + \Delta t \sum_{i=1}^{s} b_i k_i,
\end{equation}
with $\tilde{u}(t + \Delta t)$ being the approximated solution at the next time step.

Explicit methods correspond to a strictly lower-diago\-nal matrix $A$, hence avoiding any implicit dependencies.
Otherwise, the method is implicit, with coefficients, e.g., given by the collocation method~\cite{Hairera} that leads to a dense matrix $A$.
Details are skipped here for sake of brevity, but we point out the large challenge to cope with multiple stages depending implicitly on each other and various associated iterative algorithms~\cite{Emmett2010, Dutt1998, Minion2010a}.

Given the RK formulation and the demand for higher-order implicit time integration methods, efficient solvers are required to cope with the implicit dependencies in these equations.
To provide one example, a class of prominent solvers are based on spectral deferred corrections~\cite{Dutt1998}, which avoid the implicit dependency of multiple stages.
Here, the underlying idea is to iteratively correct an approximation of the solution, while each correction only requires evaluating forward and backward Euler time steps.
However, the underlying procedure is still iterative, requiring multiple iterations to gain higher-order accuracy, and hence is computationally demanding.
Quantum computers could provide a new approach to solve such problems efficiently as a combinatorial problem, as we will elaborate in Section~\ref{sec:qa_time_integration}.

\section{Digital Quantum Circuit Time Integration}
\label{sec:dqc_time_integration}

In this section, we investigate time integration by utilizing qubit-based arithmetic operations on digital quantum computers. Realizing integer operations via quantum circuits is well-known in the literature~\cite{Vedral1996, draper2000addition}. Our contribution is a generalization to fixed-point number representations.

\subsection{Arithmetic on Digital Quantum Computers}
\label{sec:arithmetic_digital_quantum_computer}

Guiding principles for the following identities can be gleaned from analogies to continuous variable quantum computing~\cite{Braunstein2005, Weedbrook2012, lloyd1999quantum2}. In this framework, a quantum register $\ket{x}$ stores a real number $x \in \R$.
This register is acted on by the position operator $\hat{\mathcal{X}}$, formally defined as
\begin{equation}
    \hat{\mathcal{X}} = \int_{\R} x \ket{x}\!\bra{x} \ud x,
\end{equation}
and its conjugate momentum operator $\hat{\mathcal{P}}$. They obey the canonical commutation relations $[\hat{\mathcal{X}}, \hat{\mathcal{P}}] = \imag \hbar$ and are related via $\hat{\mathcal{P}} = \hat{\mathcal{F}}^{\dagger} \hat{\mathcal{X}} \hat{\mathcal{F}}$, where $\hat{\mathcal{F}}$ is the Fourier transformation.

The conjugated variables can be used to realize addition and subtraction~\cite{lloyd1999quantum}.
Applying the Hamiltonian $\hat{H} = \hat{\mathcal{P}}$ for a time $c$ leads to a shift in the variable by $c$, since (setting $\hbar = 1$)
\begin{align}
    &\frac{\ud}{\ud t} \hat{\mathcal{X}} = - \imag \big[\hat{\mathcal{X}}, \hat{\mathcal{P}}\big] = 1, \\
    &\hat{\mathcal{X}} \rightarrow \hat{\mathcal{X}} + c.
\end{align}
Thus the addition can be expressed as time evolution.
Given a continuous variable quantum state $\ket{a}$, $a \in \R$, we get
\begin{equation}\label{eq:cv_addition}
    e^{-\imag c \hat{\mathcal{P}}} \ket{a} = \ket{a + c}.
\end{equation}

Our goal is to approximate this system using $n$ qubits, i.e., $2^n$ available states.
As a first step, we describe integer arithmetic and then investigate the required modifications for a fixed-point representation.
Given $a \in \{ 0, \dots, 2^n - 1 \}$, the quantum register state $\ket{a}$ is canonically identified by the corresponding tensor product of single qubit states based on the binary representation of $a$, i.e., $\ket{a} = \ket{a_{n-1}} \cdots \ket{a_1} \ket{a_0}$ for $a = a_{n-1} \cdots a_1 a_0$.
We will encounter the discrete Fourier transform at several occasions, for which we use the convention
\begin{equation}
    \hat{F} \ket{j} = \frac{1}{\sqrt{2^n}} \sum_{k=0}^{2^n-1} \e^{-2 \pi \imag j k / 2^n} \ket{k} \quad \forall j = 0, \dots, 2^n-1.
\end{equation}
$\hat{F}$ and $\hat{F}^{\dagger}$ can be efficiently implemented using the quantum Fourier transform (QFT) algorithm.

With these preparations, we define a discretized position operator by
\begin{equation}
    \hat{X} = \sum_{j=0}^{2^n-1} j \ket{j}\bra{j}.
\end{equation}
The discrete momentum operator is related to $\hat{X}$ by
\begin{equation}\label{eq:P_FXF}
    \hat{P} = \hat{F}^{\dagger} \hat{X} \hat{F},
\end{equation}
analogous to the continuous variable case.

Eq.~\eqref{eq:cv_addition} holds analogously for discrete registers~\cite{verdon2018universal}: given integers $a$ and $c$,
\begin{equation}\label{eq:add_constant}
    e^{-2 \pi \imag c \hat{P} / 2^n} \ket{a} = \ket{a + c \!\!\mod 2^n}.
\end{equation}
To realize this operation on a digital quantum computer, first note that, based on Eq.~\eqref{eq:P_FXF},
\begin{equation}\label{eq:conjugated_discrete_bases}
    e^{-2 \pi \imag c \hat{P} / 2^n} = \hat{F}^{\dagger} e^{-2 \pi \imag c \hat{X} / 2^n} \hat{F}.
\end{equation}
$e^{-2 \pi i c \hat{X} / 2^n}$ can be evaluated as follows, using that $\hat{X}$ is a diagonal matrix with respect to the computational basis:
\begin{equation}
    \begin{split}
    & e^{-2 \pi \imag c \hat{X} / 2^n} \\
    &= \sum_{a=0}^{2^n-1} e^{- 2 \pi \imag c a / 2^n} \ket{a}\bra{a} \\
    &= \sum_{a=0}^{2^n-1} \left( \prod_{k=0}^{n-1} e^{- 2 \pi \imag c a_k 2^k / 2^n} \right) \ket{a}\bra{a} \\
    &= \bigotimes_{k=0}^{n-1} \left[ \sum_{a_k \in \{0, 1\}} e^{- 2 \pi \imag c a_k / 2^{n-k}} \ket{a_k} \bra{a_k} \right].
    \end{split}
\end{equation}
The unitary matrix (qubit gate) representation of the inner operation is $R_{n-k}(c)^\dagger$, with the definition
\begin{equation}
\label{eq:phase_rotation_gate}
    R_m(\vartheta) =
    \begin{pmatrix}
    1 & 0 \\
    0 & \e^{2 \pi \imag \vartheta / 2^m}
\end{pmatrix}
\end{equation}
for any integer $m \ge 0$ and $\vartheta \in \R$. In the following, we will abbreviate $R_m(1)$ by $R_m$.

\begin{figure}[!ht]
    \centering
    \input{addition_circuit.tikz}
    \caption{An addition circuit for two quantum registers with $n = 2$, implementing Eq.~\eqref{eq:expXP_addition}; also compare with \cite{draper2000addition}.}
    \label{fig:addition_circuit}
\end{figure}
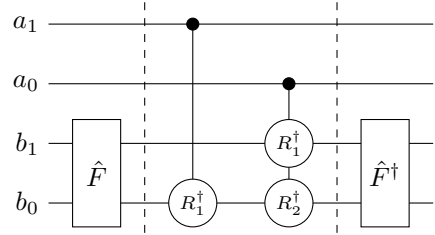

Assume we want to add two integers $a, b \in \{ 0, \dots, 2^n-1 \}$, which are both stored in quantum registers, i.e., the initial state is $\ket{a, b}$: according to Verdon et al.~\cite{verdon2018universal}, a ``von Neumann measurement'' of $\ket{a}$ combined with \eqref{eq:add_constant} can be used to increment the second register by $a$:
\begin{equation}\label{eq:expXP_addition}
    e^{- 2 \pi i \hat{X}\otimes\hat{P} / 2^n} \ket{a, b} = \ket{a, a + b \!\! \mod 2^n}.
\end{equation}
We can decompose the operator on the left as
\begin{equation}\label{eq:expXP_diagonalization}
    e^{- 2 \pi \imag \hat{X}\otimes\hat{P} / 2^n} =
    \big( I \otimes \hat{F}^{\dagger} \big) e^{- 2 \pi \imag \hat{X}\otimes\hat{X} / 2^n} \big( I \otimes \hat{F} \big),
\end{equation}
and rewrite the inner operator as
\begin{equation}\label{eq:expXX}
    \begin{split}
    &e^{- 2 \pi \imag \hat{X}\otimes\hat{X} / 2^n} = \\
    &\sum_{a=0}^{2^n-1} \sum_{b=0}^{2^n-1} \left( \prod_{j=0}^{n-1} \prod_{k=0}^{n-1-j} e^{- 2 \pi \imag a_j b_k 2^{j + k} / 2^n} \right) \ket{a, b} \bra{a, b}.
    \end{split}
\end{equation}
From this representation, one observes that $R_{m}$ gates controlled by $\ket{a}$ and applied to the $\ket{b}$ register in case $a_j \neq 0$, $b_k \neq 0$ are sufficient; see Fig.~\ref{fig:addition_circuit} for an illustration.

For general $n$, the operator in Eq.~\eqref{eq:expXX} requires $\frac{1}{2} n (n+1)$ controlled-$R_m$ gates.
Since the number of gates for the quantum Fourier transform has an $\mathcal{O}(n^2)$ scaling as well, the overall cost for the addition in Eq.~\eqref{eq:expXP_addition} is $\mathcal{O}(n^2)$ two-qubit gates (controlled phase rotations).

Analogous to the addition circuits, subtraction can be realized by letting $c \to -c \!\!\mod 2^n$, and noting that
\begin{equation}
    \begin{split}
    e^{2 \pi i \hat{X}\otimes\hat{P} / 2^n} \ket{a, b} = \ket{a, a - b \!\!\mod 2^n}
    \end{split}
\end{equation}
similar to Eq.~\eqref{eq:expXP_addition}.
Thus it suffices to take the adjoint of Eqs.~\eqref{eq:expXP_diagonalization} and \eqref{eq:expXX} to implement subtraction.

As a remark, the principle behind \eqref{eq:expXP_addition} can be generalized as follows: given a map $g: \{ 0, \dots, 2^n-1 \} \to \{ 0, \dots, 2^n-1 \}$, set
\begin{equation}
    \hat{Y}_g = \sum_{j=0}^{2^n-1} g(j) \ket{j}\bra{j}.
\end{equation}
Then
\begin{equation}
    e^{- 2 \pi i \hat{Y}_g \otimes \hat{P} / 2^n} \ket{a, b} = \ket{a, g(a) + b \!\! \mod 2^n}.
\end{equation}

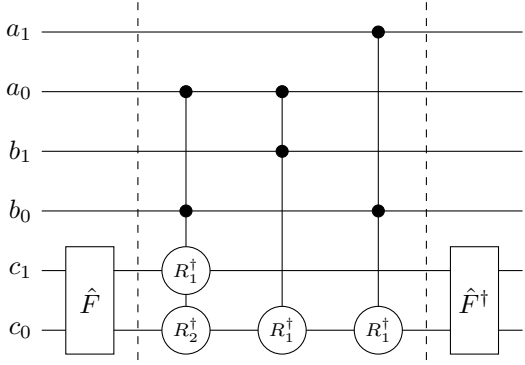
\begin{figure}[!ht]
    \centering
    \input{multiplication_circuit.tikz}
    \caption{An circuit for multiplying two integers ($n = 2$ digits), basis-encoded in states $\ket{a}$ and $\ket{b}$, and adding the result to a third qubit register $\ket{c}$; see Eq.~\eqref{eq:expXXP_multiplication}.}
    \label{fig:muliplication_circuit}
\end{figure}

The final operation we discuss here is multiplication: given $a, b, c \in \{ 0, \dots, 2^n-1 \}$, it holds that
\begin{equation}\label{eq:expXXP_multiplication}
    e^{- 2 \pi i \hat{X}\otimes\hat{X}\otimes\hat{P} / 2^n} \ket{a, b, c} = \ket{a, b, a b + c \!\! \mod 2^n},
\end{equation}
again analogous to Eq.~\eqref{eq:expXP_addition}.
We can decompose
\begin{equation}
    \begin{split}
    & e^{- 2 \pi \imag \hat{X}\otimes\hat{X}\otimes\hat{P} / 2^n} \\
    &= \big(I \otimes I \otimes \hat{F}^{\dagger}\big) e^{- 2 \pi \imag \hat{X}\otimes\hat{X}\otimes\hat{X} / 2^n} \big(I \otimes I \otimes \hat{F}^{}\big),
    \end{split}
\end{equation}
and represent the inner operator as
\begin{equation}
    \begin{split}\label{eq:multiplication_diag}
    &e^{- 2 \pi \imag \hat{X}\otimes\hat{X}\otimes\hat{X} / 2^n} = \\
    &\sum_{a,b,c = 0}^{2^n-1} \prod_{j,k,\ell=0}^{2^n-1} e^{- 2 \pi \imag a_j b_k c_\ell 2^{j+k+\ell} / 2^n} \ket{a, b, c} \bra{a, b, c}.
\end{split}
\end{equation}
All the exponential functions with $j+k+\ell \geq n$, $a_j = 0$, $b_k = 0$, or $c_\ell = 0$ evaluate to $1$ and need not be taken into account explicitly.
It suffices to act with $R_m$ gates defined in Eq.~\eqref{eq:phase_rotation_gate} on the $\ket{c}$ register, controlled by both $\ket{a}$ and $\ket{b}$; Fig.~\ref{fig:muliplication_circuit} illustrates this procedure for $n = 2$.

The overall number of doubly-controlled $R_m$ gates is $\frac{1}{6} n (2 + 3 n + n^2)$, when taking the condition $j + k + \ell < n$ into account.
This $\mathcal{O}(n^3)$ scaling, together with the required doubly-controlled gates, which have to be emulated by single and two qubit gates on current digital quantum computers, pose a considerable practical limitation on the present approach.

The advertised fixed-point representation with scaling factor $2^{-q}$ for an integer $q \ge 0$ amounts to re-interpreting a quantum state $\ket{a}$ as representing the number $2^{-q} a$, written in binary representation as
\begin{equation}
    2^{-q} a = a_{n-1} \cdots a_{q} \,.\, a_{q-1} \cdots a_0
\end{equation}
($q$ digits after the dot).
The two's complement of binary numbers is suitable for including negative numbers as well.
Then the quantum state $\ket{a}$ for $a \in \{ 0, \dots, 2^n - 1 \}$ represents
\begin{equation}
    2^{-q} \left(-2^{n-1} a_{n-1} + 2^{n-2} a_{n-2} + \dots + a_0\right)
\end{equation}
on the logical level.
The representable numbers are thus
\begin{equation}
    D_{n, q} = \left\{ \frac{-2^{n-1}}{2^q}, \frac{-2^{n-1}+1}{2^q}, \dots, \frac{2^{n-1}-1}{2^q} \right\},
\end{equation}
and the arithmetic operations are understood modulo $2^{n-q}$.
It is important to note that the operators and circuits for addition and subtraction derived so far remain exactly the same.

Concerning multiplication, the exact product of two numbers from $D_{n, q}$ requires (in general) $2 q$ digits after the dot in binary format.
To cast this back into an element of $D_{n, q}$, one disregards the trailing $q$ digits, i.e., rounds the number.
In binary representation, a bit shift to the right by $q$ places realizes this operation for non-negative numbers.
To implement this procedure using quantum circuits, we first introduce the following operator (acting on two $n$-qubit registers)
\begin{equation}
    M_q = \sum_{a,b = 0}^{2^n-1} (a b \gg q) \ket{a, b} \bra{a, b},
\end{equation}
where $\gg$ denotes the (logical) right bit shift.
Then 
\begin{equation}\label{eq:op_multiplication_fixedpoint}
    e^{- 2 \pi i M_q\otimes\hat{P} / 2^n} \ket{a, b, c} = \ket{a, b, (a b \gg q) + c \!\! \mod 2^n}.
\end{equation}
As before, we can diagonalize the operator by Fourier transformation applied to the $\ket{c}$ register.
The resulting diagonal operator then reads (compare with Eq.~\eqref{eq:multiplication_diag}):
\begin{equation}
    \begin{split}
    &e^{- 2 \pi \imag M_q\otimes\hat{X} / 2^n} = \\
    &\sum_{a,b,c = 0}^{2^n-1} \prod_{\substack{j,k,\ell=0 \\ j+k \ge q}}^{2^n-1} e^{- 2 \pi \imag a_j b_k c_\ell 2^{j+k-q+\ell} / 2^n} \ket{a, b, c} \bra{a, b, c}.
\end{split}
\end{equation}
The bit shift corresponds to the $2^{-q}$ factor in the exponent, and the rounding to the condition $j + k \ge q$.

Finally, we describe how integer division by $2$ and subsequent rounding towards $-\infty$ is achievable as quantum circuit using the two's complement representation. First note that
\begin{equation}
\begin{split}
&\frac{1}{2} \left(-2^{n-1} a_{n-1} + 2^{n-2} a_{n-2} + \dots + a_0\right) \\
&= -2^{n-1} a_{n-1} + 2^{n-2} a_{n-1} + 2^{n-3} a_{n-2} + \dots + 2^{-1} a_0.
\end{split}
\end{equation}
$a_{n-1}$ now appears twice, and rounding corresponds to dropping the term $2^{-1} a_0$.
The quantum circuit shown in Fig.~\ref{fig:divide_by_two} realizes this procedure, illustrated for $n = 4$.
\begin{figure}[!ht]
    \centering
    \input{divide_by_2.tikz}
    \caption{A circuit for dividing a basis encoded number in two's complement format by two and rounding towards $-\infty$.}
    \label{fig:divide_by_two}
\end{figure}

\subsection{Demonstration and Results}
\label{sec:qc_based_explicit_rk_time_integration}

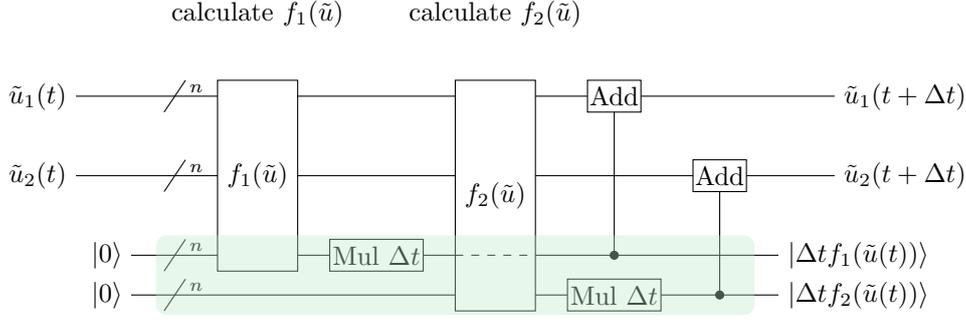
\begin{figure*}[!ht]
	\centering
	\input{explicit_euler.tikz}
	\caption{Circuit implementation of the explicit Euler scheme for Eq.~\eqref{eq:model}. The input of the circuit are two quantum registers, initialized with the values of $\tilde{u}_1(t)$ and $\tilde{u}_2(t)$ in fixed-point arithmetic. The circuit calculates one iteration of the explicit Euler scheme and outputs $\tilde{u}_1(t + \Delta t)$ and $\tilde{u}_2(t + \Delta t)$. The green region marks ancilla registers.}
	\label{fig:explicit_euler_circuit}
\end{figure*}

We employ the explicit Euler scheme as a demonstration for digital quantum circuit time integration.
In principle, the approach also works for other higher-order explicit Runge-Kutta schemes since solely elementary arithmetic operations are required, and these operations can be directly mapped to quantum gates.

Specifically, we consider the coupled linear differential equations ($N = 2$ dimensions)
\begin{equation}
    \label{eq:model}
    \frac{\ud}{\ud t} \begin{pmatrix} u_1(t) \\
    u_2(t) \end{pmatrix} = f(u(t)) := \begin{pmatrix} u_2(t) \\
    -u_1(t) \end{pmatrix},
\end{equation}
with $t \ge 0$, $u_0 = (x_1, x_2)$.
We use this system as a representation of a semi-discrete hyperbolic PDE.
The analytical solution of this equation is given by
\begin{equation}
u(t) = \begin{pmatrix}
\cos(t) & \sin(t) \\
- \sin(t) & \cos(t)
\end{pmatrix}
\begin{pmatrix}
u_1(0) \\ u_2(0)
\end{pmatrix}.
\end{equation}

To implement Eq.~\eqref{eq:model}, we need a circuit with quantum registers initialized to $\tilde{u}_1(t)$ and $\tilde{u}_2(t)$.
The output of the circuit are registers storing $\tilde{u}_1(t + \Delta t)$ and $\tilde{u}_2(t + \Delta t)$.
We also need two ``ancilla'' temporary registers for evaluating $f(\tilde{u}(t))$.
The overall circuit is depicted in Fig.~\ref{fig:explicit_euler_circuit}.
Note that applying a sequence of time steps requires a reinitialization of the ancilla registers to $\ket{0}$; alternatively, this could be achieved by ``uncomputing'' them.

To implement the circuit, $\Delta t$ has to be a power of $\frac{1}{2}$, which allows us to save qubits during the simulation.
This is solely for efficiency reasons and other $\Delta t$ values could be realized as well.
Instead of multiplying by $\Delta t$, we can divide multiple times by $2$.
Note, that $\Delta t$ does not need to be stored in a quantum register.

\begin{figure}[!htb]
    \centering
    \subfloat[$f_1(u) = u_2$]{\input{f_1_dot.tikz}}\\
    \subfloat[$f_2(u) = - u_1$]{\input{f_2_dot.tikz}}
    \caption{Quantum circuits implementing $f(u)$ for the model problem in Eq.~\eqref{eq:model}.}
    \label{fig:f_circuit}
\end{figure}
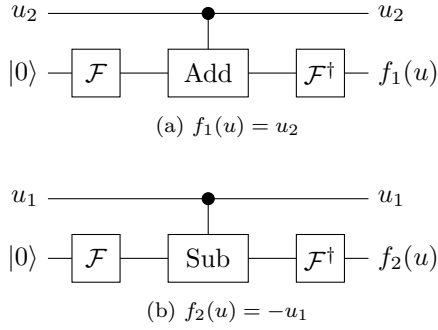
Fig.~\ref{fig:f_circuit} shows the circuits implementing the right side $f(u)$ of the differential equation \eqref{eq:model}.

We simulate the quantum circuits using Qiskit \cite{Qiskit}, without noise or decoherence.
The time step is chosen as $\Delta t = \frac{1}{2}$.
Each number register consists of $4$ qubits, and the fixed-point arithmetic scaling factor is set to $q = 1$.

As verification, we implement an explicit Euler also on a classical computer and compare the results.
In all cases, the results match in a numerical sense, as expected.

\begin{figure}[!htb]
    \centering
    \subfloat[stationary equilibrium point $u(t) = (0, 0)$]{\includegraphics[width=0.86\linewidth]{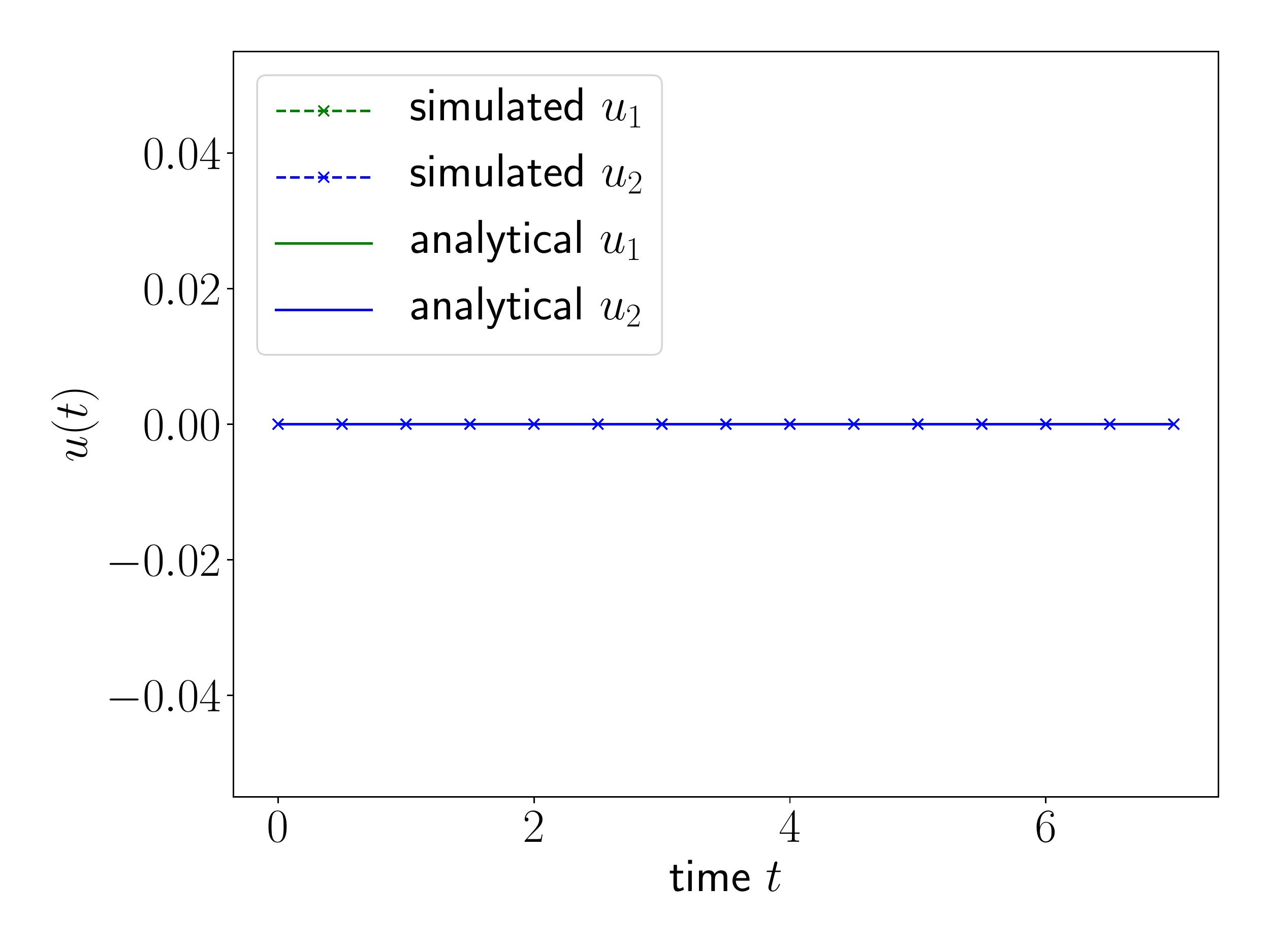}\label{fig:explicit_euler_equilibrium}}\\
    \subfloat[initial value $u_0 = (0, -1)$]{\includegraphics[width=0.86\linewidth]{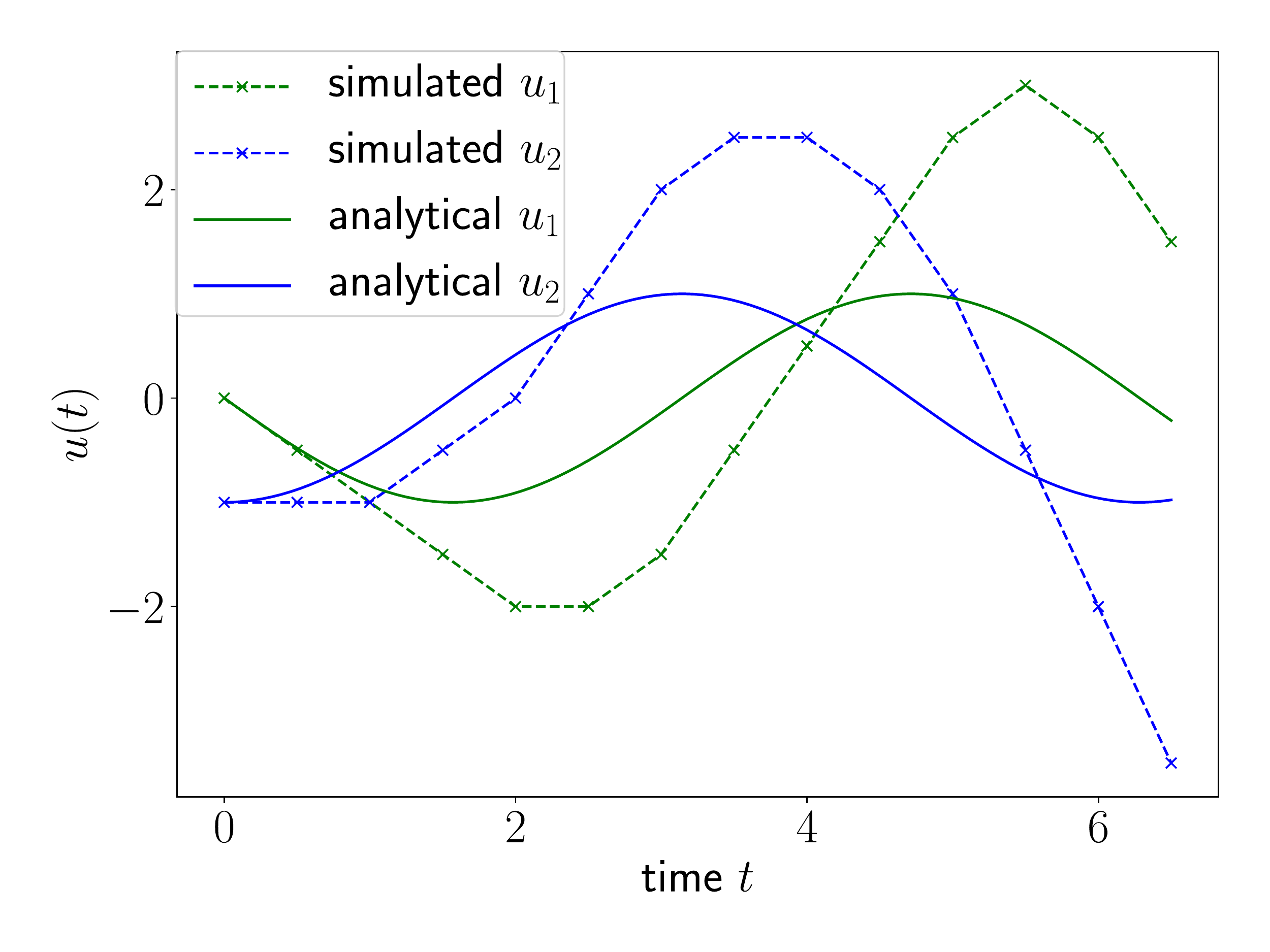}\label{fig:explicit_euler_excited}}
    \caption{Explicit Euler method for the linear ODEs \eqref{eq:model}, implemented on a digital quantum computer. The simulated solution is represented by dashed lines, and the analytical solution by solid lines. The large error originates from the low-order time integration method and large time step size.}
\end{figure}

We first study the stationary case ($\ud u/\ud t = 0$) at the equilibrium point $u(t) = (0, 0)$, see Fig.~\ref{fig:explicit_euler_equilibrium}.
Although this seems to be trivial, such stationary cases play an important role for, e.g., more complex cases like geostrophic balance in atmospheric simulations\cite{Staniforth2012}, and therefore are included here as proof-of-concept.
Indeed the DQC solution remains constant as well, as expected.

Next, we simulate the same differential equation, but using initial values $u_0 = (0, -1)$; the results are visualized in Fig.~\ref{fig:explicit_euler_excited}.
The simulated solution amplifies the real one, as expected for the explicit Euler scheme~\cite{Durran2010}:
since the digital quantum computer mimics the operations of a classical solver, the achievable accuracy is likewise bounded by the numerical errors of the classical method. Our demonstrations mainly serve as proof-of-principle.

\section{Quantum Annealing Time Integration}
\label{sec:qa_time_integration}

Annealing is a method for solving optimization problems.
The solution of a problem is encoded into the global minimum (``ground state'') of an energy function.
One possible encoding, relevant for combinatorial optimization problems, is quadratic unconstrained binary optimization (QUBO), see, e.g., Glover et al.~\cite{glover2018tutorial}.
In this formulation, the goal is to minimize the binary variables $\sigma_i \in \{0, 1\}$ of the energy function (Ising-type Hamiltonian)
\begin{equation}
\hat{H} = - \sum_{i,j} J_{ij} \sigma_i \sigma_j - h \sum_i \sigma_i,
\end{equation}
for given parameters $J_{i,j}, h_i \in \R$.
While thermal annealing uses thermal fluctuations to overcome local minima of the target function, quantum annealing uses quantum tunneling effects for that purpose, potentially exhibiting a faster convergence \cite{kadowaki1998quantum}.
Nevertheless, neither thermal nor quantum annealing are guaranteed to find the ground state of the optimization problem.

\subsection{Time Integration as an Optimization Problem}

In the following, we consider a system of autonomous linear differential equations
\begin{equation}\label{eq:linear_dgl}
\frac{\ud}{\ud t} u(t) = f(u(t)) := \sum_{i=1}^{M} L^{(i)} u(t)^{\otimes (i-1)}
\end{equation}
with $u(t) \in \R^N$ for $t \ge 0$, $u(0) = u_0$, $L^{(i)} \in \left(\R^{N}\right)^{\otimes i}$, and $M$ is the highest polynomial degree of the differential equation.
An extension to other nonlinear equations is possible with our framework, as well, as long as the action of $f$ can be formulated as part of a QUBO problem.

A time step of collocation-based implicit Runge-Kutta time integration~\cite{Hairera} can be represented in the form of
\begin{equation}\label{eq:collocationRK_repr}
    \tilde{u}(t + \Delta t) = \tilde{u}(t) + \Delta t S f\big(\tilde{u}(t + \Delta t)\big)
\end{equation}
with $\tilde{u}(0)$ a suitably chosen vector with the initial conditions, and $S$ the spectral integration matrix given by the particular quadrature method used in the collocation formulation.
In this formulation, the vector $\tilde{u}$ also stores all intermediate state solutions at the quadrature points.
We can interpret \eqref{eq:collocationRK_repr} as an optimization problem (see, e.g., \cite{Emmett2010}) of the form
\begin{equation}
    \tilde{u}(t + \Delta t) = \argmin_{v \in \R^N} \lVert v - \Delta t S f(v) - \tilde{u}(t)\rVert.
\end{equation}

In the following, we will phrase this optimization problem in the context of a quantum annealer, to obtain $\tilde{u}(t+\Delta t)$ for a given $\tilde{u}(t)$.
A similar approach for solving linear systems of equations has already been introduced by O'Malley and Vesselinov in~\cite{malley2016}.
Our approach for the extension to polynomial equations is mostly inspired by Cheng Chang et al~\cite{Chang2019quantum}.
Based on the Runge-Kutta methods (see Eqs.~\eqref{eq:rk_stage_equation} and~\eqref{eq:rk_closure_equation}) and using Eq.~\eqref{eq:linear_dgl}, we derive
\begin{align}\label{eq:runge-kutta1}
    &\tilde{u}_j(t + \Delta t) = \tilde{u}_j(t) + \Delta t \sum_{o = 1}^{s} b_o K_{oj},
\end{align}
\begin{align}\label{eq:runge-kutta2}
    &K_{oj} = \sum_{\substack{i=1 \\  k_1,...,k_i \in [N]}}^{M} L^{(i)}_{jk_1...k_i} \tilde{u}_{k_1}(t) ... \tilde{u}_{k_i}(t) +  \notag \\
    &\Delta t \sum_{e=1}^{s} A_{oe} \sum_{\substack{i=1 \\  k_1,...,k_i \in [N]}}^{M} L^{(i)}_{jk_1...k_i} K_{e k_1} ... K_{e k_i},
\end{align}
with $K \in \R^{s \times N}$ and $[N] = \{1, ..., N\}$.
In order to solve this equation on a quantum annealer, we rewrite it as an optimization problem with real variables and then reformulate that as a QUBO. 
First, taking the squared difference of Eq.~\eqref{eq:runge-kutta1} and Eq.~\eqref{eq:runge-kutta2} leads to the minimization problem (for time points $t_m = m \, \Delta t$):
\begin{widetext}
\begin{equation} \label{eq:minimization_problem}
\begin{split}
\tilde{u}_j(t_{m+1})
&= \argmin_{\tilde{u}_j(t_{m+1})} \min_{K_1, \dots, K_s} \left( \tilde{u}_j(t_{m+1}) 
- \tilde{u}_j(t_m) - \Delta t  \sum_{o = 1}^{s} b_o K_{oj} \right)^2 \\
&+ \sum_{o=1}^{s} \left( K_{oj} - \sum_{\substack{i=1 \\  k_1,...,k_i \in [N]}}^{M} L^{(i)}_{jk_1...k_i} \tilde{u}_{k_1}(t_m) ... \tilde{u}_{k_i}(t_m)
 - \Delta t \sum_{e=1}^{s} A_{oe} \sum_{\substack{i=1 \\  k_1,...,k_i \in [N]}}^{M} L^{(i)}_{jk_1...k_i} K_{e k_1} ... K_{e k_i} \right)^2 
%
\end{split}
\end{equation}
\end{widetext}
The variables to optimize are $\tilde{u}_j(t_{m+1})$ and $K_{oj}$ for $o = 1, \dots, s$, $j = 1, \dots, N$.
The vector $\tilde{u}_j(t_m)$ remains constant in Eq.~\eqref{eq:minimization_problem} and, therefore, is not part of the optimization.
Unfortunately, we cannot directly map this equation to the Ising model because the parameters to optimize are real.
To overcome this issue, we develop (another) tailored approximation using binary variables, as shown below.

\subsection{Binary Number Representation}

Different from the fixed-point representation described in Section~\ref{sec:arithmetic_digital_quantum_computer}, here our goal is an adaptive version (similar to the floating-point format).
We approximate a real number $g \in \R$ by
\begin{equation}\label{eq:number_representation_annealer}
    \begin{split}
    g \approx \Phi_{d}^{k}(\sigma)
    &= 2^{n-1-k} \sigma_{n-1} + \dots + 2^{0-k} \sigma_{0} + d \\
    &= 2^{-k} \sigma + d,
    \end{split}
\end{equation}
where $d, k \in \R$ are parameters and $\sigma_i$ (for $i = 0, \dots, n-1$) the available binary variables.
During an optimization step on a quantum computer, only the binary variables $(\sigma_i)$ are modified.
The detailed procedure will be described below.

\subsection{Variational Approach}

\label{sec:variational_circuit_integration_annealing}
One limitation of current quantum annealers is the connectivity between qubits available in hardware.
In order to use arbitrary couplings in the QUBO model, embeddings have to be used.
An embedding maps a QUBO problem onto a slightly different one. 
This new model solves the same task using fewer connections per qubit, but a larger overall number of qubits~\cite{Lechner2015}.
Thus, in general, an important goal is the design of algorithms with sparse connectivity and small qubit counts.
Typically, one chooses a variational approach to achieve this~\cite{McClean_2016}.
The key idea is to only run a part of the problem on the quantum device and perform post-processing on a classical computer to optimize parameters.
This procedure is repeated until the output converges to the final solution.
However, convergence using the post-processing technique is only guaranteed in the convex case, which in general is only satisfied for linear actions $f$.

In the following, we analyze the number of qubits and connectivity required by our algorithm to achieve a given numerical accuracy.
Subsequently, we introduce a variational formulation.

\subsubsection{Numerical Accuracy and Qubit Connectivity}

We assume that $n$ qubits are sufficient to approximate the range of values needed to represent an entry of the solution vector.
By choosing $k = n$ and $d = 0$ in Eq.~\eqref{eq:number_representation_annealer}, this interval is $[0, 1]$, and the accuracy for each simulated number is given by $2^{-n}$.
Furthermore, we assume that the solution of the DEQ remains in this interval.

The overall number of required qubits for $s$ stages in the Runge-Kutta method is $n (s + 1)$, since an additional $n$ qubits store the result $\tilde{u}_j(t + \Delta t)$ at the next time step.
Additionally, auxiliary qubits are needed to reduce cubic and higher terms to quadratic ones. 
To achieve this, we use reduction by substitution which was first presented by Rosenberg~\cite{rosenberg1975reduction}.
For this method, one auxiliary qubit per reduction is used.
The amount of auxiliary qubits rises exponentially with the order $M$, but is also limited by the number of stages $s$ and the system size of the DEQ.
For the remainder of this analysis, we will only assume linear DEQ ($M = 1$) for which no auxiliary qubits are needed.

Next, we determine the required connectivity between qubits. According to Eq.~\eqref{eq:minimization_problem}, a multiplication between two registers requires $n$ connections between the qubits and a multiplication with itself $n-1$ qubits.
Eq.~\eqref{eq:minimization_problem} also shows that the connectivity depends both on the density of the Runge-Kutta table $A$, as well as the DEQ matrix $L$.
For our analysis, we consider the worst-case scenario of dense matrices.

There are three sets of terms in Eq.~\eqref{eq:minimization_problem} in which two different registers are multiplied. 
The first term is the multiplication of $\tilde{u}_j(t_{m+1})$ with itself (directly after the second equal sign in \eqref{eq:minimization_problem}), for which $n-1$ connections are needed.
In the second set of terms, the entries of $K$ are multiplied with each other (several occasions after the second equal sign). Since there are $s N$ of these entries, where $N$ is the system size, we count $s N (s N - 1) / 2$ multiplications in total.
However, $s N$ of these multiplications are with the same registers.
Therefore, there are $s N (s N - 1) / 2 - s N = s^2 N^2 / 2 - 3 s N / 2$ multiplications with other and $s N$ multiplications with the same register.
As a consequence, the total number of connections needed for the second scenario is $s^2 N^2 n / 2 - 3 s N n / 2 + s N (n - 1) = s^2 N^2 n / 2 - s N n/2 - s N$.
In the third set of terms, all entries of $K$ are multiplied with the entries of $\tilde{u}_j(t_{m+1})$. In total, these are $s N$ multiplications, accounting for $s N n$ connections.
All of the three sets of terms create new connections, and therefore the connection counts must be summed up, leading to a total number of connections required for this algorithm (in the worst case) of
\begin{equation}\label{eq:connectivity}
    \sharp(\text{connect.}) = n-1 + s^2 N^2 n / 2 + s N n/2 - s N.
\end{equation}

\subsubsection{Variational Adaptive Number Representation}

According to the $s^2 N^2 n / 2$ term in Eq.~\eqref{eq:connectivity}, increasing the number of qubits $n$ for approximating a real number will demand a much higher connectivity on the quantum device.
To mitigate this issue, we switch to an adaptive number representation, as already anticipated in Eq.~\eqref{eq:number_representation_annealer}.
Specifically, for each individual number $g$ appearing in the algorithm, we iteratively refine $k$ and $d$ in Eq.~\eqref{eq:number_representation_annealer}, denoted $g_i$, $k_i$ and $d_i$ for the $i$\textsuperscript{th} iteration.
As update step, we set
\begin{align}
k_{i+1} &= k_i + c, \label{eq:ki_update} \\
d_{i+1} &= g_i - 2^{n-k_{i+1}-1} \label{eq:di_update}
\end{align}
with a fixed constant $c > 0$. The idea, reminiscent of nested intervals, is to shrink the range of values covered by $2^{-k} \sigma$ (which is optimized by the quantum computer) in Eq.~\eqref{eq:number_representation_annealer}, while absorbing the best candidate for $g$ into the classical offset parameter $d$. 
The shift by $2^{n-k_{i+1}-1}$ in Eq.~\eqref{eq:di_update} ensures that both positive and negative corrections are possible.
We remark that, as our experiments in the next section shows, $c$ must not be chosen too large on real quantum annealers, and should ideally correspond to the convergence rate of the iteration.

\subsubsection{Time Complexity}
In the following, we analyze the time complexity of our algorithm.
Since, to our knowledge, there is neither any evidence that quantum annealers have an advantage compared to classical computers, nor we are aware of any method to determine the time complexity of an algorithm on a quantum annealer, we study the time complexity of the algorithm on an adiabatic quantum computer.
In this type of quantum computing, the solution of a problem is found by evolving a time-dependent Hamiltonian where the solution is encoded into the ground state of the final Hamiltonian.
The system is first initialized to the ground state of an initial Hamiltonian. If the evolution is slow enough, at the end of the evolution the system will remain in the state of the desired solution.
The time the evolution takes scales with $T = O\left(\frac{1}{g_{min}^2}\right)$, where $g_{min}$ is the smallest energy gap between the ground state and another state during the evolution process~\cite{farhi2000quantum}.

We choose an initial Hamiltonian $H_{init}$ with a sufficient high energy gap, so that it does not limit the time complexity of our algorithm. 
The time-dependent Hamiltonian we use is given by
\begin{align}
	\hat{H}(t) = \left(1-\frac{t}{T}\right) \hat{H}_{\mathrm{init}} +  \frac{t}{T} \hat{H}_{\mathrm{QUBO}},
\end{align}
where $\hat{H}_{\mathrm{QUBO}}$ is given by the QUBO Hamiltonian of our minimization function.
Although the real minimal gap energy gap $g_{\mathrm{min}}$ is found by determining the energy difference of the ground and first excited state for all $\hat{H}(t) \quad \forall t$, we determine this gap for the QUBO Hamiltonian $\hat{H}_{\mathrm{QUBO}}$ in order to get an idea how this gap changes with parameters we chose in the algorithm.
Afterwards, we compare our findings with numerical results of the minimal gap in $\hat{H}(t)$.

For linear ODEs the QUBO Hamiltonian is strictly convex. This makes it possible to calculate boundaries on the energy gap under the assumption that the lowest energy is $0$.
We first show the calculation in detail for the explicit Euler of a one dimensional ODE and later deal with the generalized Runge-Kutta case of linear ODEs.
For the explicit Euler of a linear ODE with $L^{(2)} = \Lambda$, a time step is given as 
\begin{align}
	u(t+\Delta t) = u(t) + \Delta t \Lambda u(t).
\end{align}
In our Hamiltonian, $u(t)$ is exactly represented, while $u(t + \Delta t)$ is discretized and represented by different quantum states. 
Therefore, we use the notation $u(t + \Delta t) = u_{n+1} \Delta x$, where $u_{n+1} \in \mathbb{Z}$ and $\Delta x$ is the smallest number resolution in our algorithm.
The solution for $u_{n+1}$ which minimizes Eq.~\eqref{eq:minimization_problem} is given by
\begin{align}
	u^*_{n+1} = \frac{u(t) + \Delta t \Lambda u(t)}{\Delta x}.
\end{align}
We assume that $u^*_{n+1} \in \mathbb{Z}$, which means that a state for the energy of $0$ exists.
Because of the construction, we know that this is the lowest energy state and that the minimization function is symmetric.
Therefore, the second lowest energy level is given by
\begin{align}
	u^{\epsilon}_{n+1} = u^*_{n+1} \pm 1.
\end{align}
To find the energy gap $g_{min}$, we can simply subtract the energy of $u^{\epsilon}_{n+1}$ by the energy of $u^*_{n+1}$.
Our finding is that
\begin{align}
	g_{\mathrm{min}} = \Delta x^2.
\end{align}
Next, we compare this result to the scaling of the minimal gap in $\hat{H}(t)$, which we computed numerically using and initial state of $\hat{H}_{\mathrm{init}} = - 0.5 \left(1 - \sigma_x\right) \otimes ... \otimes 0.5 \left(1 - \sigma_x\right)$.
\begin{figure}[!ht]
	\centering
	\includegraphics[width=\linewidth]{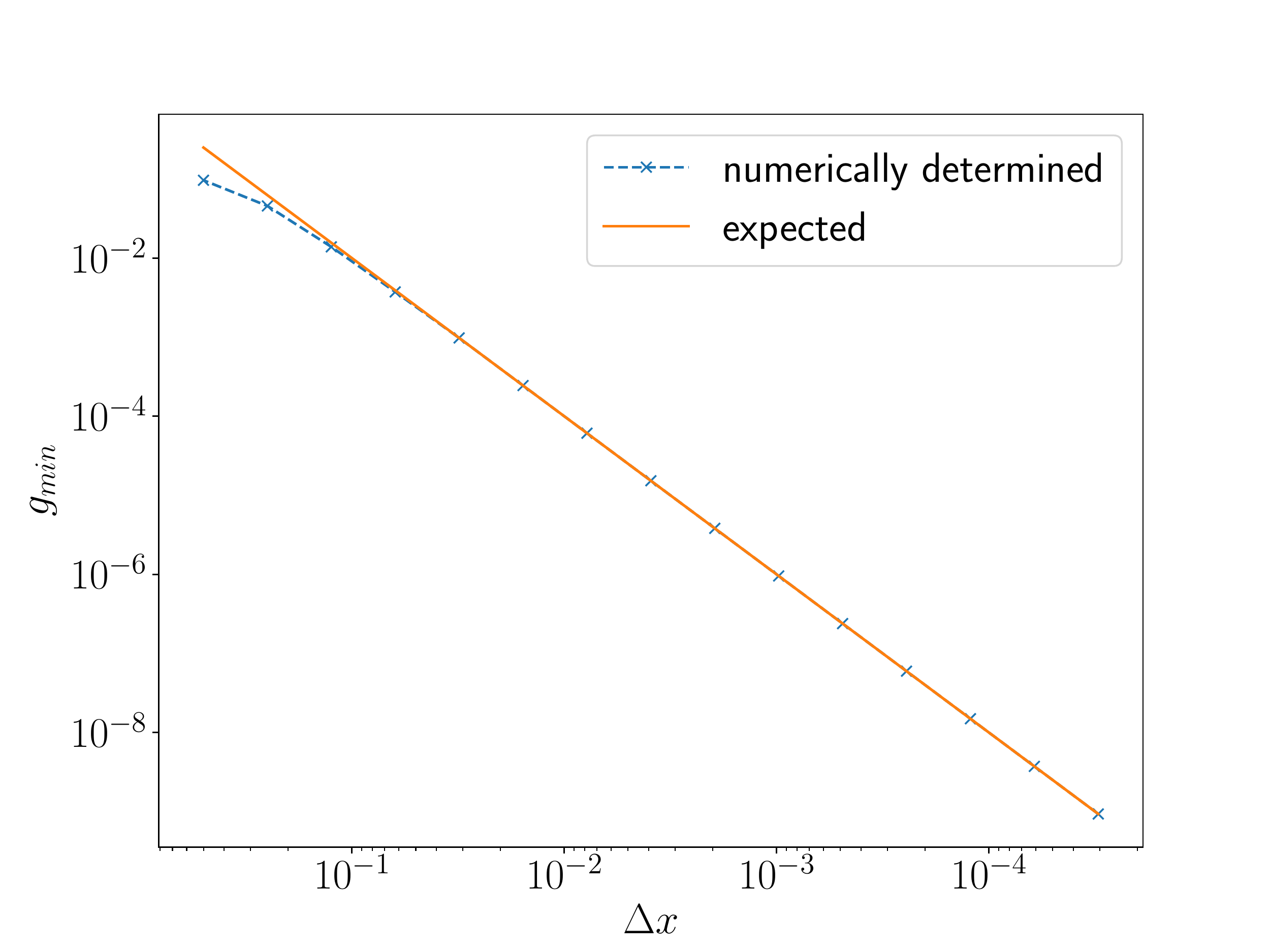}
	\caption{Comparison of the scaling of the energy gap between the ground and first excited state for the adiabatic Hamiltonian for the explicit Euler. For sufficient small $\Delta x$, the expected scaling and numerical determined scaling match.}
	\label{fig:scaling_explicit_euler_dx}
\end{figure}
The comparison is depicted in Fig.~\ref{fig:scaling_explicit_euler_dx}. For sufficient small $\Delta x$ our previously determined scaling and numerical results perfectly match.
Another important parameter to choose is $\Delta t$. According to our derivation for $\hat{H}_{\mathrm{QUBO}}$, a change in $\Delta t$ should not have any influence in $g_{\mathrm{min}}$.
The comparison of the numerical and expected results confirms that changes in $\Delta t$ do not have an influence on $g_{\mathrm{min}}$.
Our assumption that the scaling of the energy gap of the ground and first excited state of $\hat{H}_{\mathrm{QUBO}}$ is comparable to the overall minimal energy gap $g_{\mathrm{min}}$ agrees with numerical results.
Therefore, we conclude that the time complexity of the explicit Euler using our algorithm approximately is
\begin{align}
	T \approx O\left(\frac{1}{\Delta x^4}\right).
\end{align}

Next, we determine the complexity of a generalized Runge-Kutta method.
In this case, we have to discretize the stages as well, $K_o = K_{o,n} \Delta x$, where $K_{o, n} \in \mathbb{Z}$.
We refer to the minimizer of Eq.~\eqref{eq:minimization_problem} using the $*$ symbol and assume that these are contained in $\mathbb{Z}$.
Let us define the deviation from a solution $K_{o, n}$ and $u_{n+1}$ to these minimizers as $\Delta K_{o, n} = \left| K_{o, n} - K_{o, n}^* \right|$ and $\Delta u_{n+1} = \left| u_{n+1} - u_{n+1}^* \right|$.
With these definitions, the minimal energy gap in the QUBO Hamiltonian is calculated by
\begin{align}\label{eq:minimization_for_gap}
	g_{\mathrm{QUBO}} &= \min \left(\Delta u_{n+1} \Delta x - \Delta t \sum_{o=1}^{s} b_o \Delta K_{o, n} \Delta x\right)^2  \notag \\
	&+ \sum_{o=1}^{s} \left(\Delta K_{o, n} \Delta x - \Delta t \sum_{e=1}^{s} A_{oe} \Lambda \Delta K_{e, n} \Delta x\right)^2,
\end{align}
where $\Delta u_{n+1} + \sum_{o=1}^{s} \Delta K_{o,n} \geq 1$.
First, note that $\Delta u_{n+1}, \Delta K_{1, n}, ... , \Delta K_{s, n} \in \{0, 1\}$ in order to minimize Eq.~\eqref{eq:minimization_for_gap}.
In the following, we want to give a lower bound on the energy gap using only the first term of Eq.~\eqref{eq:minimization_for_gap}.
Under the assumption, that $\Delta t < 1$, this term is either minimized if all variables deviate or only a single of the stages. Therefore,
\begin{align}\label{eq:boundary_g_final}
	g_{\mathrm{QUBO}} \geq \min \left\{\Delta x^2 \left(1 - \Delta t \right)^2, \Delta t^2 \Delta x^2 b^2 \right\},
\end{align}
where $b = \min\{b_i \text{ for } 1 \leq i \leq s\}$. 
We verify that the bound also holds for $\hat{H}(t)$ by evaluating the minimal gap numerically. As a Runge-Kutta scheme, we use Crank-Nicolson.
\begin{figure}[!ht]
	\centering
	\includegraphics[width=\linewidth]{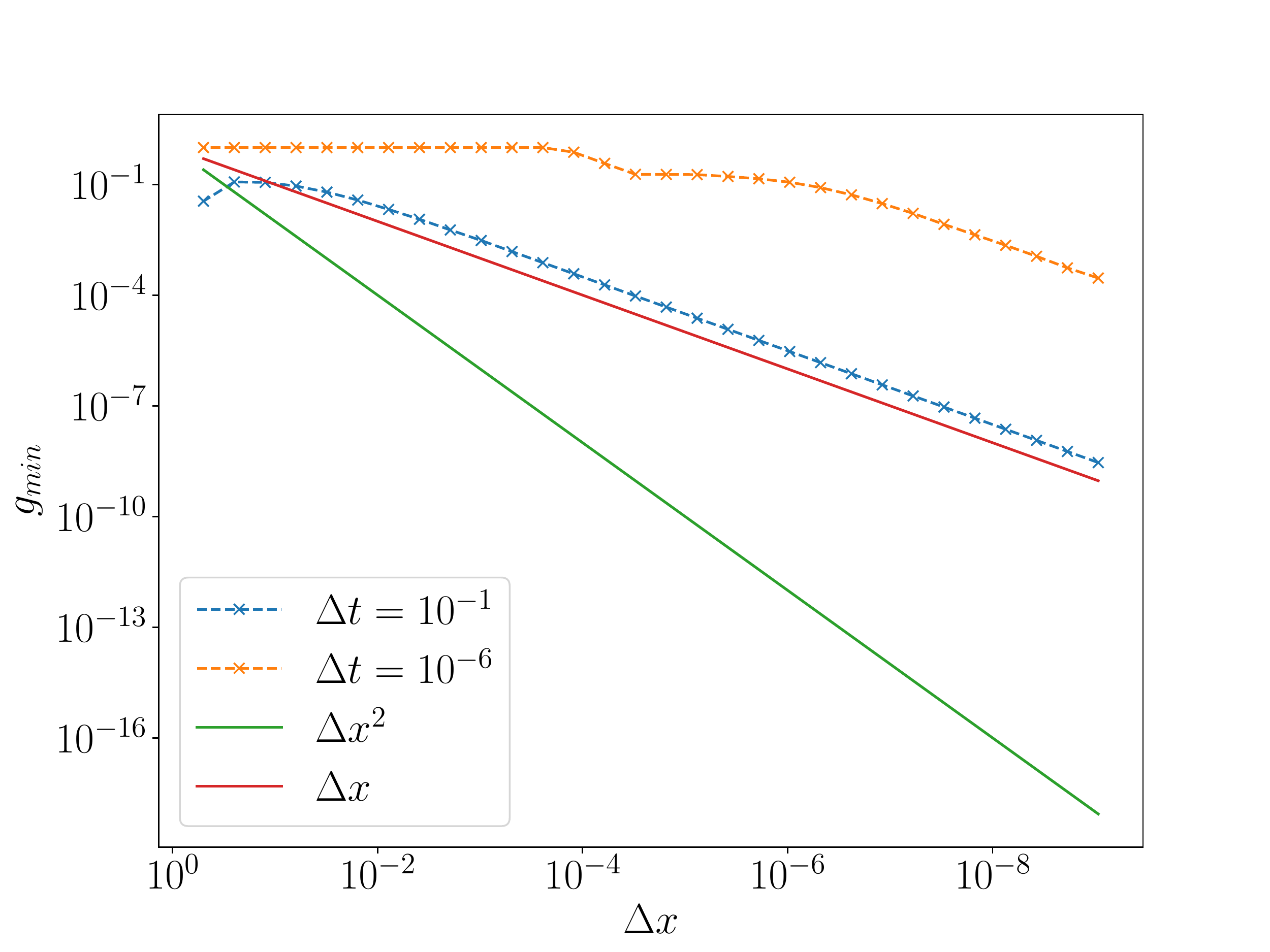}
	\caption{Comparison of the scaling of the energy gap between the ground and first excited state for the adiabatic Hamiltonian for the Crank-Nicolson scheme with changing $\Delta x$. It can be seen that $g_{\mathrm{min}}$ is proportional to $\Delta x$ and higher time steps result in a lower energy gap.}
	\label{fig:scaling_crank-nicolson_dx}
\end{figure}
The scaling in $\Delta x$ for $\Delta t = 0.1$ and $\Delta t = 10^{-6}$ is depicted in Fig.~\ref{fig:scaling_crank-nicolson_dx}.
For this specific case, it can be seen that the minimal gap has a better scaling, $g_{\mathrm{min}} \propto O \left(\Delta x\right)$, compared to the lower bound on $g_{\mathrm{QUBO}}$ we found.
For our concrete example, we can also see that higher time steps result in a lower energy gap.

To get a scaling including the amount of stages, we limit the Runge-Kutta scheme on the collocation method using Chebyshev-Gauss quadrature in the following.
This allows $b = 1/s$ for increasing amount of stages.
From the boundary we found in Eq.~\eqref{eq:boundary_g_final}, we conclude that the time complexity of the Runge-Kutta method, using Chebyshev-Gauss collocation, is approximately better than
\begin{align}
	T \lesssim O\left(\max \left\{\frac{s^4}{\Delta x^4 \Delta t^4}, \frac{1}{\Delta x^4 \left(1 - \Delta t\right)^4} \right\}\right).
\end{align}

For both the explicit Euler case, as well as the generalized Runge-Kutta case, it is possible to generalize the results to systems of equations of arbitrary sizes.
This is since the term we used in the minimization function is independent of any mutual interference for other variables in a system of equations.
Compared to a classical computer, where so far no parallel processing without an additional time scaling is known, we don't have that in this quantum algorithm.
Instead, a larger system of equations only allocates more hardware/qubit resources.

\subsection{Results of the QA Time Integration}
\label{subsec:discussion_qa_time_integration}

We perform the quantum annealing task on the D-Wave 2000Q system, which 
 features 2041 qubits.
To encode the QUBO model onto the annealer, the values of the $J$ matrix and the $h$ vector are first normalized to be within the interval $[-2, 1]$ for the $J$ matrix and $[-2, 2]$ for the $h$ vector.
For our problem, we directly map the values of $J$ and $h$ onto the interval $\left[-1, 1\right]$.
Furthermore, as arbitrary connectivity between all qubits is not possible, the concrete connections available on the target annealing hardware have to be taken into account and the QUBO model has to be adapted to those hardware characteristics.
If the QUBO problem needs more connections than the hardware provides, an additional embedding has to be found.
We solve both of these tasks using the library \emph{minorminer}, which uses a method given by Cai et al.~\cite{cai2014practical}.
We perform $100$ reads on the annealer per iteration. 
For the last example, we additionally used the library \emph{PyQUBO} \cite{tanahashi2019application} 
which maps higher order polynomials to quadratic ones using the reduction by substitution method.

\begin{figure}[!ht]
    \centering
    \includegraphics[width=\linewidth]{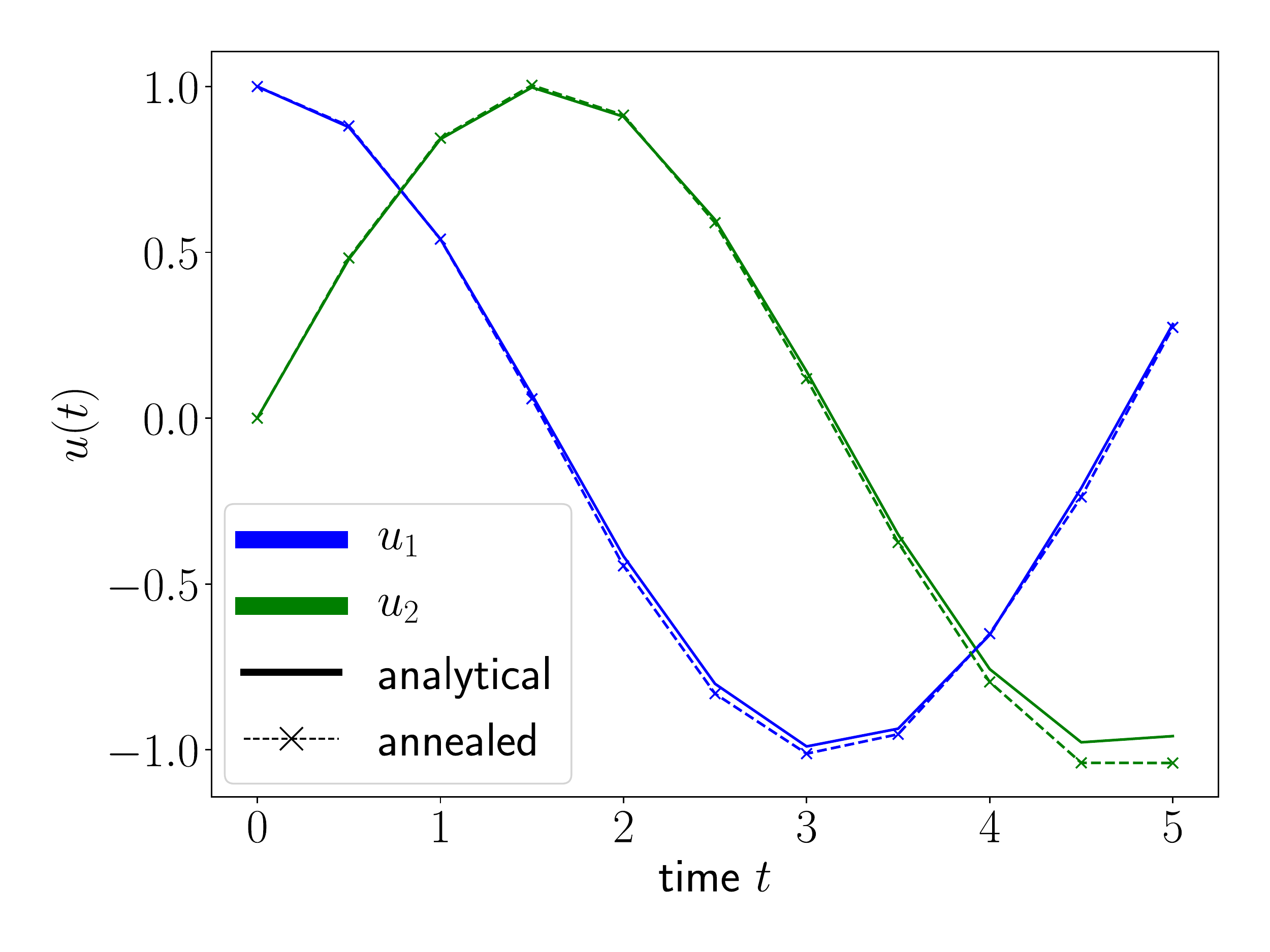}
    \caption{Comparison of the analytical (exact) and annealed result of Eq.~\eqref{eq:model} using the 6\textsuperscript{th} order Gauss-Legendre collocation method with a timestep $\Delta t = 0.5$. The numerical result is visually indistinguishable from the analytical (exact) solution.}
    \label{fig:example_annealing_10lambda}
\end{figure}
Fig.~\ref{fig:example_annealing_10lambda} shows the analytical and annealed result of the linear coupled differential equation given in Eq.~\eqref{eq:model} with the initial conditions $u_1(0) = 1$ and $u_2(0) = 0$.
As a specific integration method, we choose the Gauss-Legendre collocation method of order $6$ with a timestep of $\Delta t = 0.5$.
For every grid point, $3$ qubits are used.
We perform $15$ iterations per timestep, starting with $k_0 = 1$ and $c = 0.5$.
The numerical result obtained on a digital computer, using the same numerical method as for the annealing result, is visually indistinguishable from the exact solution.
Therefore, the solution of the quantum annealer does not perfectly match the numerical solution of a classical computer.
This is both due to errors in the quantum hardware as well as not enough iterations per timestep in our algorithm.
We expect that the errors of future quantum annealers will be significantly lower.

Next, we analyze the convergence of our variational approach by performing a single Runge-Kutta step on a quantum annealer and compare it to numerical results using the same Runge-Kutta method.
\begin{figure}[!ht]
	\centering
	\includegraphics[width=\linewidth]{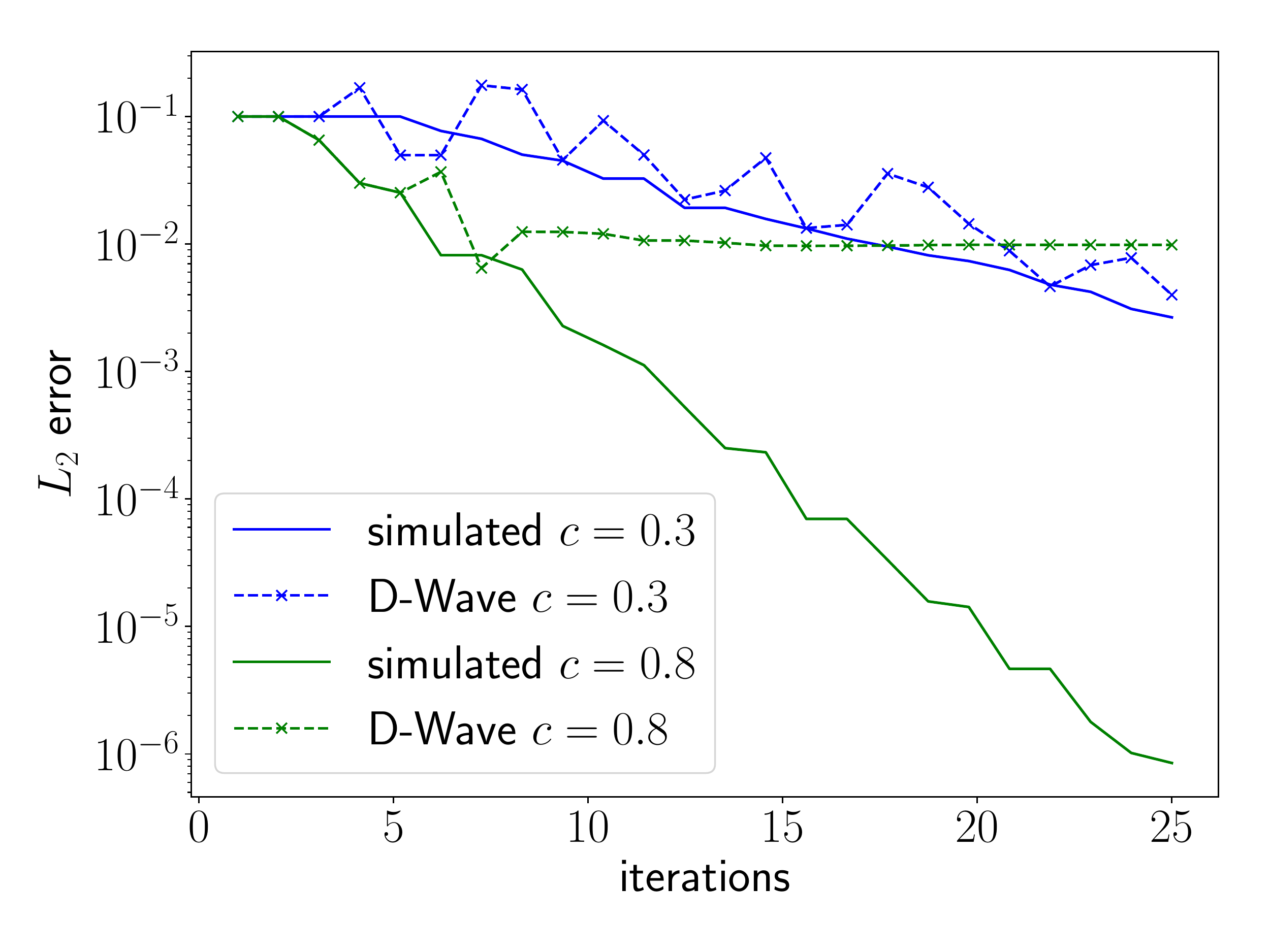}
	\caption{Error of the annealed Gauss-Legendre 6\textsuperscript{th} order Runge-Kutta step using our variational approach, compared to a numerical evaluation of the same Runge-Kutta step.
	The dashed lines are obtained on a D-Wave quantum annealer, and the solid lines show an exact annealing reference solution calculated on a classical computer.
	The parameter $c$ is the exponent shift of the variational approach in Eq.~\eqref{eq:ki_update}. For all cases, two qubits per number register are used.}
	\label{fig:variational_convergence}
\end{figure}
Fig.~\ref{fig:variational_convergence} shows these errors for the Gauss-Legendre collocation method of order $6$.
The dashed line results are obtained on the D-Wave 2000Q system, while the solid lines are simulated on a digital computer by an exact solver, which finds the exact minimum of the combinatorial optimization function.
We use two different exponent shift constants, given in Eq.~\eqref{eq:ki_update}, for the variational approach:
The blue lines use $c = 0.3$, the green lines use $c=0.8$.
For the higher exponent shift, $c=0.8$, first, the annealing solution follows the exact solution closely and converges faster than the lower convergence case. Nevertheless, after an error spike at the tenth iteration, the convergence stops and the error of the annealing solution is not further reduced.
For $c=0.3$, the results of the quantum annealer follows the exact solver over all iterations, except of some spikes, which are caused by errors in the computing hardware. However, due to the lower exponent shift, the variational approach is still able to recover the same result even with errors present.

Last, we solve a Ricatti equation, given by
\begin{align}
	\frac{\ud}{\ud t} u(t) = u(t) - u(t)^2
\end{align}
using our algorithm. 
This DEQ has two equilibrium points, an unstable one at $u = 0$ and a stable one at $u = 1$.
As the integration method we used the Crank-Nicolson method with $2$ qubits per register.
For this case, only two auxiliary qubits are needed,
making it $8$ qubits in total, excluding those qubits which are needed for the embedding on a quantum annealer.
Although the variational approach is not guaranteed to converge, we try the variational approach with $c = 0.8$ and $10$ iterations.
For this nonlinear case, we only present the results from an annealing simulator, not on the actual D-Wave 2000Q system.
This is because of the non convex structure of the problem and high noise on the D-Wave system the results are hard to interpret.
First, we tested the unstable equilibrium point and verified that the solution stays at the unstable equilibrium point, $u = 0$.
Due to the high noise on current quantum annealers, this result is not achieved on a physical device.
However, it shows that our algorithm is able to evaluate such solutions correctly.
\begin{figure}[!ht]
	\centering
	\includegraphics[width=\linewidth]{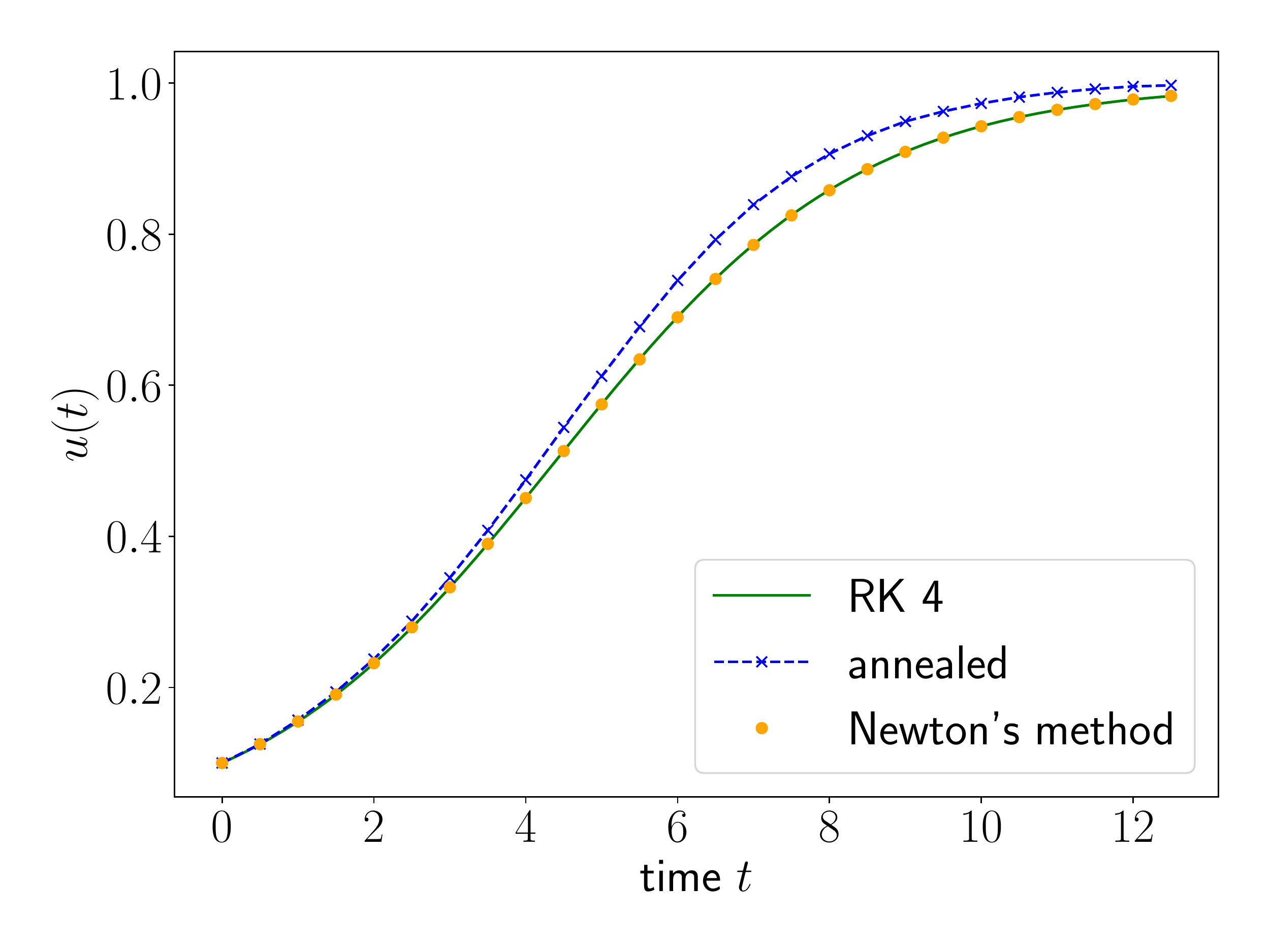}
	\caption{Comparison of the simulated annealed, analytical, and Newton method for $u(0) = 0.1$. The analytical solution was obtained by using the 4\textsuperscript{th} Runge-Kutta method with much smaller time step sizes. The simulated annealed solution does not match the result of the Newton method.}
	\label{fig:nonlinear_dynamic}
\end{figure}
Next, we evaluate a dynamic case, starting at $u(0) = 0.1$.
In this case, the solution is expected to converge to the stable point $u(\infty) = 1$.
The results of the exact annealing are depicted in Fig.~\ref{fig:nonlinear_dynamic}.
Using the same integration method and time step, the annealed solution does not match the Newton method.
We verified that the results by the Newton's algorithm are global minima of Eq.~\eqref{eq:minimization_problem}.
The variational approach does not guarantee to converge to a global minima for non convex problems.
Without the variational approach, with a sufficient amount of qubits for each register, and Eq.~\eqref{eq:minimization_problem} being continuous and only having a single global minimum, we believe that our annealing algorithm is superior to the Newton's method.
Namely, Newton's method is likely prone to local minima if the initial estimate for the method is not chosen correctly.
A sufficient amount of qubits in our annealing algorithm ensures that a configuration close to the global minimum is a better minimizer to Eq.~\eqref{eq:minimization_problem} compared to any other configuration close to a local minimum.
However, this assumes hardware which is free of noise and a large amount of qubits.

Our findings serve as proof-of-concept and demonstrates the feasibility of our algorithm on current quantum annealers.
The second example demonstrates that the exponent shift parameter has to be chosen wisely, dependent on the error of the quantum hardware.
Finding best suitable exponent shifts for specific hardware is not covered in this paper and additional research is needed.
The third example shows that with our algorithm the Runge-Kutta method can also be applied to nonlinear ODEs.
However, for the nonlinear example studied, the annealed results do not match the expected results.
A discussion about conditions, which we believe are necessary for nonlinear ODEs, is provided.

\section{Discussion and Outlook}
\label{sec:summary_and_conclusion}

We have explored the usage of quantum computers for the task of solving differential equations, a ubiquitous problem in engineering and scientific computing.
Our focus lies on the technique instead of a benchmark comparison, as the latter is not (yet) reasonable given the small problem size that had to studied here.

Concerning the time integration based on arithmetic operations on a digital quantum computer (Section~\ref{sec:dqc_time_integration}), one might ask how this could possibly provide an advantage, given that the operations of a classical computer are directly mimicked.
As scenario for exploiting the inherent parallelism of quantum computers, the circuits developed here could be used as building blocks within encompassing quantum algorithms, e.g., as an ``oracle'' within a quantum search method.
Specifically, in the context of ordinary differential equations, this approach could solve inverse problems directly, like finding initial conditions given a desired state at the final time.
From a broader perspective, we imagine that quantum computers could serve as parallel accelerators of subroutines within classical algorithms (like time integration).

Looking at the quantum annealing approach (Section~\ref{sec:qa_time_integration}), we see the largest potential for high-order implicit time integration methods, as the associated computational costs scales more favorably with integration order when using quantum annealing (compared to classical computers).
In addition, we were able to decouple the number of required qubits from the accuracy of the number representation for the quantum annealing approach, and the arithmetic steps are implemented implicitly into the QUBO model, reducing the qubit count for temporary variables.
Even though this decoupling is only guaranteed to work for linear ODEs, we applied it to a nonlinear ODE where our results did not match the expected results.
We provided a discussion about the conditions, which we believe are necessary to solve nonlinear ODEs with our algorithm.
We also gave an upper bound for the time complexity of our annealing algorithm if it is applied to an adiabatic quantum computer in the linear case. 
We showed that this complexity does not depend on the size of the problem itself, but rather on the discretization size of the variables and the time step size.

As an interesting alternative direction, we point out that the inherent dynamics of quantum systems could be used for solving nonlinear differential equations via the Koopman-von Neumann formulation of classical mechanics~\cite{Joseph2020}.

\begin{acknowledgments}
We would like to thank Udo Helmbrecht and Wolfgang Gehrke from the research institute CODE at the Universit\"at der Bundeswehr M\"unchen for facilitating a preliminary exploration of arithmetic operations on an ``ibmq\_toronto'' digital quantum computer via IBM Q cloud access. We acknowledge support from the Munich Quantum Center (MQC) and the TUM Institute for Advanced Study (TUM-IAS).
This project has received funding from the Federal Ministry of Education and Research and the European High-Performance Computing Joint Undertaking (JU) under grant agreement No 955701. The JU receives support from the European Union's Horizon 2020 research and innovation programme and Belgium, France, Germany, Switzerland.
\end{acknowledgments}

\end{document}

%% file: addition_circuit.tikz
\begin{tikzpicture}[scale=1.500000,x=1pt,y=1pt]
\filldraw[color=white] (0.000000, -7.500000) rectangle (96.000000, 52.500000);
\draw[color=black] (0.000000,45.000000) -- (96.000000,45.000000);
\draw[color=black] (0.000000,45.000000) node[left] {$a_1$};
\draw[color=black] (0.000000,30.000000) -- (96.000000,30.000000);
\draw[color=black] (0.000000,30.000000) node[left] {$a_0$};
\draw[color=black] (0.000000,15.000000) -- (96.000000,15.000000);
\draw[color=black] (0.000000,15.000000) node[left] {$b_1$};
\draw[color=black] (0.000000,0.000000) -- (96.000000,0.000000);
\draw[color=black] (0.000000,0.000000) node[left] {$b_0$};
\draw (12.000000,15.000000) -- (12.000000,0.000000);
\begin{scope}
\draw[fill=white] (12.000000, 7.500000) +(-45.000000:8.485281pt and 19.091883pt) -- +(45.000000:8.485281pt and 19.091883pt) -- +(135.000000:8.485281pt and 19.091883pt) -- +(225.000000:8.485281pt and 19.091883pt) -- cycle;
\clip (12.000000, 7.500000) +(-45.000000:8.485281pt and 19.091883pt) -- +(45.000000:8.485281pt and 19.091883pt) -- +(135.000000:8.485281pt and 19.091883pt) -- +(225.000000:8.485281pt and 19.091883pt) -- cycle;
\draw (12.000000, 7.500000) node {{$\hat{F}$}};
\end{scope}
\draw (36.000000,45.000000) -- (36.000000,0.000000);
\begin{scope}
\draw[fill=white] (36.000000, 0.000000) circle(6.000000pt);
\clip (36.000000, 0.000000) circle(6.000000pt);
\draw (36.000000, 0.000000) node {{\scriptsize $R_1^{\dagger}$}};
\end{scope}
\filldraw (36.000000, 45.000000) circle(1.500000pt);
\draw (60.000000,30.000000) -- (60.000000,0.000000);
\begin{scope}
\draw[fill=white] (60.000000, 0.000000) circle(6.000000pt);
\clip (60.000000, 0.000000) circle(6.000000pt);
\draw (60.000000, 0.000000) node {{\scriptsize $R_2^{\dagger}$}};
\end{scope}
\begin{scope}
\draw[fill=white] (60.000000, 15.000000) circle(6.000000pt);
\clip (60.000000, 15.000000) circle(6.000000pt);
\draw (60.000000, 15.000000) node {{\scriptsize $R_1^{\dagger}$}};
\end{scope}
\filldraw (60.000000, 30.000000) circle(1.500000pt);
\draw (84.000000,15.000000) -- (84.000000,0.000000);
\begin{scope}
\draw[fill=white] (84.000000, 7.500000) +(-45.000000:8.485281pt and 19.091883pt) -- +(45.000000:8.485281pt and 19.091883pt) -- +(135.000000:8.485281pt and 19.091883pt) -- +(225.000000:8.485281pt and 19.091883pt) -- cycle;
\clip (84.000000, 7.500000) +(-45.000000:8.485281pt and 19.091883pt) -- +(45.000000:8.485281pt and 19.091883pt) -- +(135.000000:8.485281pt and 19.091883pt) -- +(225.000000:8.485281pt and 19.091883pt) -- cycle;
\draw (84.000000, 7.500000) node {{$\hat{F}^{\dagger}$}};
\end{scope}
\draw[dashed] (24.000000, -7.500000) -- (24.000000, 52.500000);
\draw[dashed] (72.000000, -7.500000) -- (72.000000, 52.500000);
\end{tikzpicture}

%% file: multiplication_circuit.tikz
\begin{tikzpicture}[scale=1.500000,x=1pt,y=1pt]
\filldraw[color=white] (0.000000, -7.500000) rectangle (120.000000, 82.500000);
\draw[color=black] (0.000000,75.000000) -- (120.000000,75.000000);
\draw[color=black] (0.000000,75.000000) node[left] {$a_1$};
\draw[color=black] (0.000000,60.000000) -- (120.000000,60.000000);
\draw[color=black] (0.000000,60.000000) node[left] {$a_0$};
\draw[color=black] (0.000000,45.000000) -- (120.000000,45.000000);
\draw[color=black] (0.000000,45.000000) node[left] {$b_1$};
\draw[color=black] (0.000000,30.000000) -- (120.000000,30.000000);
\draw[color=black] (0.000000,30.000000) node[left] {$b_0$};
\draw[color=black] (0.000000,15.000000) -- (120.000000,15.000000);
\draw[color=black] (0.000000,15.000000) node[left] {$c_1$};
\draw[color=black] (0.000000,0.000000) -- (120.000000,0.000000);
\draw[color=black] (0.000000,0.000000) node[left] {$c_0$};
\draw (12.000000,15.000000) -- (12.000000,0.000000);
\begin{scope}
\draw[fill=white] (12.000000, 7.500000) +(-45.000000:8.485281pt and 19.091883pt) -- +(45.000000:8.485281pt and 19.091883pt) -- +(135.000000:8.485281pt and 19.091883pt) -- +(225.000000:8.485281pt and 19.091883pt) -- cycle;
\clip (12.000000, 7.500000) +(-45.000000:8.485281pt and 19.091883pt) -- +(45.000000:8.485281pt and 19.091883pt) -- +(135.000000:8.485281pt and 19.091883pt) -- +(225.000000:8.485281pt and 19.091883pt) -- cycle;
\draw (12.000000, 7.500000) node {{$\hat{F}$}};
\end{scope}
\draw (36.000000,60.000000) -- (36.000000,0.000000);
\begin{scope}
\draw[fill=white] (36.000000, 0.000000) circle(6.000000pt);
\clip (36.000000, 0.000000) circle(6.000000pt);
\draw (36.000000, 0.000000) node {{\scriptsize $R_2^{\dagger}$}};
\end{scope}
\begin{scope}
\draw[fill=white] (36.000000, 15.000000) circle(6.000000pt);
\clip (36.000000, 15.000000) circle(6.000000pt);
\draw (36.000000, 15.000000) node {{\scriptsize $R_1^{\dagger}$}};
\end{scope}
\filldraw (36.000000, 60.000000) circle(1.500000pt);
\filldraw (36.000000, 30.000000) circle(1.500000pt);
\draw (60.000000,60.000000) -- (60.000000,0.000000);
\begin{scope}
\draw[fill=white] (60.000000, 0.000000) circle(6.000000pt);
\clip (60.000000, 0.000000) circle(6.000000pt);
\draw (60.000000, 0.000000) node {{\scriptsize $R_1^{\dagger}$}};
\end{scope}
\filldraw (60.000000, 60.000000) circle(1.500000pt);
\filldraw (60.000000, 45.000000) circle(1.500000pt);
\draw (84.000000,75.000000) -- (84.000000,0.000000);
\begin{scope}
\draw[fill=white] (84.000000, 0.000000) circle(6.000000pt);
\clip (84.000000, 0.000000) circle(6.000000pt);
\draw (84.000000, 0.000000) node {{\scriptsize $R_1^{\dagger}$}};
\end{scope}
\filldraw (84.000000, 75.000000) circle(1.500000pt);
\filldraw (84.000000, 30.000000) circle(1.500000pt);
\draw (108.000000,15.000000) -- (108.000000,0.000000);
\begin{scope}
\draw[fill=white] (108.000000, 7.500000) +(-45.000000:8.485281pt and 19.091883pt) -- +(45.000000:8.485281pt and 19.091883pt) -- +(135.000000:8.485281pt and 19.091883pt) -- +(225.000000:8.485281pt and 19.091883pt) -- cycle;
\clip (108.000000, 7.500000) +(-45.000000:8.485281pt and 19.091883pt) -- +(45.000000:8.485281pt and 19.091883pt) -- +(135.000000:8.485281pt and 19.091883pt) -- +(225.000000:8.485281pt and 19.091883pt) -- cycle;
\draw (108.000000, 7.500000) node {{$\hat{F}^{\dagger}$}};
\end{scope}
\draw[dashed] (24.000000, -7.500000) -- (24.000000, 82.500000);
\draw[dashed] (96.000000, -7.500000) -- (96.000000, 82.500000);
\end{tikzpicture}

%% file: divide_by_2.tikz
\begin{tikzpicture}[scale=1.500000,x=1pt,y=1pt]
\filldraw[color=white] (0.000000, -7.500000) rectangle (81.000000, 52.500000);
\draw[color=black] (0.000000,45.000000) -- (81.000000,45.000000);
\draw[color=black] (0.000000,45.000000) node[left] {$a_3$};
\draw[color=black] (0.000000,30.000000) -- (81.000000,30.000000);
\draw[color=black] (0.000000,30.000000) node[left] {$a_2$};
\draw[color=black] (0.000000,15.000000) -- (81.000000,15.000000);
\draw[color=black] (0.000000,15.000000) node[left] {$a_1$};
\draw[color=black] (0.000000,0.000000) -- (81.000000,0.000000);
\draw[color=black] (0.000000,0.000000) node[left] {$a_0$};
\draw[color=black] (13.500000, 0.000000) node [fill=white] {${\ket{0}}$};
\draw (36.000000,45.000000) -- (36.000000,0.000000);
\begin{scope}
\draw[fill=white] (36.000000, 0.000000) circle(3.000000pt);
\clip (36.000000, 0.000000) circle(3.000000pt);
\draw (33.000000, 0.000000) -- (39.000000, 0.000000);
\draw (36.000000, -3.000000) -- (36.000000, 3.000000);
\end{scope}
\filldraw (36.000000, 45.000000) circle(1.500000pt);
\draw (54.000000,15.000000) -- (54.000000,0.000000);
\begin{scope}
\draw (51.878680, -2.121320) -- (56.121320, 2.121320);
\draw (51.878680, 2.121320) -- (56.121320, -2.121320);
\end{scope}
\begin{scope}
\draw (51.878680, 12.878680) -- (56.121320, 17.121320);
\draw (51.878680, 17.121320) -- (56.121320, 12.878680);
\end{scope}
\draw (72.000000,30.000000) -- (72.000000,15.000000);
\begin{scope}
\draw (69.878680, 12.878680) -- (74.121320, 17.121320);
\draw (69.878680, 17.121320) -- (74.121320, 12.878680);
\end{scope}
\begin{scope}
\draw (69.878680, 27.878680) -- (74.121320, 32.121320);
\draw (69.878680, 32.121320) -- (74.121320, 27.878680);
\end{scope}
\end{tikzpicture}

%% file: explicit_euler.tikz
\definecolor{owngreen}{RGB}{190,235,202}
\begin{tikzpicture}[scale=1.000000,x=1pt,y=1pt]
\filldraw[color=white] (0.000000, -7.500000) rectangle (284.000000, 97.500000);
\draw[color=black] (0.000000,75.000000) -- (284.000000,75.000000);
\draw[color=black] (0.000000,75.000000) node[left] {${\tilde{u}_1(t)}$};
\draw[color=black] (0.000000,45.000000) -- (284.000000,45.000000);
\draw[color=black] (0.000000,45.000000) node[left] {${\tilde{u}_2(t)}$};
\draw[color=black] (13.500000,15.000000) -- (270.500000,15.000000);
\draw[color=black] (13.500000,0.000000) -- (270.500000,0.000000);
\draw[color=black] (21.000000,15.000000) node[fill=white,left,minimum height=15.000000pt,minimum width=15.000000pt,inner sep=0pt] {\phantom{${\ket{0}}$}};
\draw[color=black] (21.000000,15.000000) node[left] {${\ket{0}}$};
\draw[color=black] (21.000000,0.000000) node[fill=white,left,minimum height=15.000000pt,minimum width=15.000000pt,inner sep=0pt] {\phantom{${\ket{0}}$}};
\draw[color=black] (21.000000,0.000000) node[left] {${\ket{0}}$};
\draw (33.000000, 69.000000) -- (41.000000, 81.000000);
\draw (39.000000, 78.000000) node[right] {$\scriptstyle{n}$};
\draw (33.000000, 39.000000) -- (41.000000, 51.000000);
\draw (39.000000, 48.000000) node[right] {$\scriptstyle{n}$};
\draw (33.000000, 9.000000) -- (41.000000, 21.000000);
\draw (39.000000, 18.000000) node[right] {$\scriptstyle{n}$};
\draw (33.000000, -6.000000) -- (41.000000, 6.000000);
\draw (39.000000, 3.000000) node[right] {$\scriptstyle{n}$};
\draw (68.000000, 97.500000) node[text width=144pt,above,text centered] {calculate $f_1(\tilde{u})$};
\draw (68.000000,75.000000) -- (68.000000,15.000000);
\begin{scope}
\draw[fill=white] (68.000000, 45.000000) +(-45.000000:21.213203pt and 50.911688pt) -- +(45.000000:21.213203pt and 50.911688pt) -- +(135.000000:21.213203pt and 50.911688pt) -- +(225.000000:21.213203pt and 50.911688pt) -- cycle;
\clip (68.000000, 45.000000) +(-45.000000:21.213203pt and 50.911688pt) -- +(45.000000:21.213203pt and 50.911688pt) -- +(135.000000:21.213203pt and 50.911688pt) -- +(225.000000:21.213203pt and 50.911688pt) -- cycle;
\draw (68.000000, 45.000000) node {{$f_1(\tilde{u})$}};
\end{scope}
\begin{scope}
\draw[fill=white] (112.500000, 15.000000) +(-45.000000:24.748737pt and 8.485281pt) -- +(45.000000:24.748737pt and 8.485281pt) -- +(135.000000:24.748737pt and 8.485281pt) -- +(225.000000:24.748737pt and 8.485281pt) -- cycle;
\clip (112.500000, 15.000000) +(-45.000000:24.748737pt and 8.485281pt) -- +(45.000000:24.748737pt and 8.485281pt) -- +(135.000000:24.748737pt and 8.485281pt) -- +(225.000000:24.748737pt and 8.485281pt) -- cycle;
\draw (112.500000, 15.000000) node {{Mul $\Delta t$}};
\end{scope}
\draw (157.000000, 97.500000) node[text width=144pt,above,text centered] {calculate $f_2(\tilde{u})$};
\draw (157.000000,75.000000) -- (157.000000,0.000000);
\begin{scope}
\draw[fill=white] (157.000000, 37.500000) +(-45.000000:21.213203pt and 61.518290pt) -- +(45.000000:21.213203pt and 61.518290pt) -- +(135.000000:21.213203pt and 61.518290pt) -- +(225.000000:21.213203pt and 61.518290pt) -- cycle;
\clip (157.000000, 37.500000) +(-45.000000:21.213203pt and 61.518290pt) -- +(45.000000:21.213203pt and 61.518290pt) -- +(135.000000:21.213203pt and 61.518290pt) -- +(225.000000:21.213203pt and 61.518290pt) -- cycle;
\draw (157.000000, 37.500000) node {{$f_2(\tilde{u})$}};
\end{scope}
\draw[color=black,dashed] (142.000000, 15.000000) -- (172.000000, 15.000000);
\draw (201.500000,75.000000) -- (201.500000,15.000000);
\begin{scope}
\draw[fill=white] (201.500000, 75.000000) +(-45.000000:14.142136pt and 8.485281pt) -- +(45.000000:14.142136pt and 8.485281pt) -- +(135.000000:14.142136pt and 8.485281pt) -- +(225.000000:14.142136pt and 8.485281pt) -- cycle;
\clip (201.500000, 75.000000) +(-45.000000:14.142136pt and 8.485281pt) -- +(45.000000:14.142136pt and 8.485281pt) -- +(135.000000:14.142136pt and 8.485281pt) -- +(225.000000:14.142136pt and 8.485281pt) -- cycle;
\draw (201.500000, 75.000000) node {{Add}};
\end{scope}
\filldraw (201.500000, 15.000000) circle(1.500000pt);
\begin{scope}
\draw[fill=white] (201.500000, 0.000000) +(-45.000000:24.748737pt and 8.485281pt) -- +(45.000000:24.748737pt and 8.485281pt) -- +(135.000000:24.748737pt and 8.485281pt) -- +(225.000000:24.748737pt and 8.485281pt) -- cycle;
\clip (201.500000, 0.000000) +(-45.000000:24.748737pt and 8.485281pt) -- +(45.000000:24.748737pt and 8.485281pt) -- +(135.000000:24.748737pt and 8.485281pt) -- +(225.000000:24.748737pt and 8.485281pt) -- cycle;
\draw (201.500000, 0.000000) node {{Mul $\Delta t$}};
\end{scope}
\draw (241.000000,45.000000) -- (241.000000,0.000000);
\begin{scope}
\draw[fill=white] (241.000000, 45.000000) +(-45.000000:14.142136pt and 8.485281pt) -- +(45.000000:14.142136pt and 8.485281pt) -- +(135.000000:14.142136pt and 8.485281pt) -- +(225.000000:14.142136pt and 8.485281pt) -- cycle;
\clip (241.000000, 45.000000) +(-45.000000:14.142136pt and 8.485281pt) -- +(45.000000:14.142136pt and 8.485281pt) -- +(135.000000:14.142136pt and 8.485281pt) -- +(225.000000:14.142136pt and 8.485281pt) -- cycle;
\draw (241.000000, 45.000000) node {{Add}};
\end{scope}
\filldraw (241.000000, 0.000000) circle(1.500000pt);
\draw[color=black] (263.000000,15.000000) node[fill=white,right,minimum height=15.000000pt,minimum width=15.000000pt,inner sep=0pt] {\phantom{${\ket{\Delta t f_1(\tilde{u}(t))}}$}};
\draw[color=black] (263.000000,15.000000) node[right] {${\ket{\Delta t f_1(\tilde{u}(t))}}$};
\draw[color=black] (263.000000,0.000000) node[fill=white,right,minimum height=15.000000pt,minimum width=15.000000pt,inner sep=0pt] {\phantom{${\ket{\Delta t f_2(\tilde{u}(t))}}$}};
\draw[color=black] (263.000000,0.000000) node[right] {${\ket{\Delta t f_2(\tilde{u}(t))}}$};
\draw[color=black] (284.000000,75.000000) node[right] {${\tilde{u}_1(t + \Delta t)}$};
\draw[color=black] (284.000000,45.000000) node[right] {${\tilde{u}_2(t + \Delta t)}$};
\draw[draw opacity=0.000000,fill opacity=0.200000,fill=owngreen,rounded corners] (30.000000,22.500000) rectangle (254.000000,-7.500000);
\draw[draw opacity=0.000000,fill opacity=0.200000,fill=owngreen,rounded corners] (30.000000,22.500000) rectangle (254.000000,-7.500000);
\end{tikzpicture}

%% file: f_1_dot.tikz
\begin{tikzpicture}[scale=1.500000,x=1pt,y=1pt]
\filldraw[color=white] (0.000000, -7.500000) rectangle (80.000000, 22.500000);
\draw[color=black] (0.000000,15.000000) -- (80.000000,15.000000);
\draw[color=black] (0.000000,15.000000) node[left] {$u_2$};
\draw[color=black] (0.000000,0.000000) -- (80.000000,0.000000);
\draw[color=black] (0.000000,0.000000) node[left] {$\ket{0}$};
\begin{scope}
\draw[fill=white] (12.000000, -0.000000) +(-45.000000:8.485281pt and 8.485281pt) -- +(45.000000:8.485281pt and 8.485281pt) -- +(135.000000:8.485281pt and 8.485281pt) -- +(225.000000:8.485281pt and 8.485281pt) -- cycle;
\clip (12.000000, -0.000000) +(-45.000000:8.485281pt and 8.485281pt) -- +(45.000000:8.485281pt and 8.485281pt) -- +(135.000000:8.485281pt and 8.485281pt) -- +(225.000000:8.485281pt and 8.485281pt) -- cycle;
\draw (12.000000, -0.000000) node {{$\mathcal{F}$}};
\end{scope}
\draw (40.000000,15.000000) -- (40.000000,0.000000);
\begin{scope}
\draw[fill=white] (40.000000, -0.000000) +(-45.000000:14.142136pt and 8.485281pt) -- +(45.000000:14.142136pt and 8.485281pt) -- +(135.000000:14.142136pt and 8.485281pt) -- +(225.000000:14.142136pt and 8.485281pt) -- cycle;
\clip (40.000000, -0.000000) +(-45.000000:14.142136pt and 8.485281pt) -- +(45.000000:14.142136pt and 8.485281pt) -- +(135.000000:14.142136pt and 8.485281pt) -- +(225.000000:14.142136pt and 8.485281pt) -- cycle;
\draw (40.000000, -0.000000) node {{Add}};
\end{scope}
\filldraw (40.000000, 15.000000) circle(1.500000pt);
\begin{scope}
\draw[fill=white] (68.000000, -0.000000) +(-45.000000:8.485281pt and 8.485281pt) -- +(45.000000:8.485281pt and 8.485281pt) -- +(135.000000:8.485281pt and 8.485281pt) -- +(225.000000:8.485281pt and 8.485281pt) -- cycle;
\clip (68.000000, -0.000000) +(-45.000000:8.485281pt and 8.485281pt) -- +(45.000000:8.485281pt and 8.485281pt) -- +(135.000000:8.485281pt and 8.485281pt) -- +(225.000000:8.485281pt and 8.485281pt) -- cycle;
\draw (68.000000, -0.000000) node {{$\mathcal{F}^{\dagger}$}};
\end{scope}
\draw[color=black] (80.000000,15.000000) node[right] {$u_2$};
\draw[color=black] (80.000000,0.000000) node[right] {$f_1(u)$};
\end{tikzpicture}

%% file: f_2_dot.tikz
\begin{tikzpicture}[scale=1.500000,x=1pt,y=1pt]
\filldraw[color=white] (0.000000, -7.500000) rectangle (80.000000, 22.500000);
\draw[color=black] (0.000000,15.000000) -- (80.000000,15.000000);
\draw[color=black] (0.000000,15.000000) node[left] {$u_1$};
\draw[color=black] (0.000000,0.000000) -- (80.000000,0.000000);
\draw[color=black] (0.000000,0.000000) node[left] {$\ket{0}$};
\begin{scope}
\draw[fill=white] (12.000000, -0.000000) +(-45.000000:8.485281pt and 8.485281pt) -- +(45.000000:8.485281pt and 8.485281pt) -- +(135.000000:8.485281pt and 8.485281pt) -- +(225.000000:8.485281pt and 8.485281pt) -- cycle;
\clip (12.000000, -0.000000) +(-45.000000:8.485281pt and 8.485281pt) -- +(45.000000:8.485281pt and 8.485281pt) -- +(135.000000:8.485281pt and 8.485281pt) -- +(225.000000:8.485281pt and 8.485281pt) -- cycle;
\draw (12.000000, -0.000000) node {{$\mathcal{F}$}};
\end{scope}
\draw (40.000000,15.000000) -- (40.000000,0.000000);
\begin{scope}
\draw[fill=white] (40.000000, -0.000000) +(-45.000000:14.142136pt and 8.485281pt) -- +(45.000000:14.142136pt and 8.485281pt) -- +(135.000000:14.142136pt and 8.485281pt) -- +(225.000000:14.142136pt and 8.485281pt) -- cycle;
\clip (40.000000, -0.000000) +(-45.000000:14.142136pt and 8.485281pt) -- +(45.000000:14.142136pt and 8.485281pt) -- +(135.000000:14.142136pt and 8.485281pt) -- +(225.000000:14.142136pt and 8.485281pt) -- cycle;
\draw (40.000000, -0.000000) node {{Sub}};
\end{scope}
\filldraw (40.000000, 15.000000) circle(1.500000pt);
\begin{scope}
\draw[fill=white] (68.000000, -0.000000) +(-45.000000:8.485281pt and 8.485281pt) -- +(45.000000:8.485281pt and 8.485281pt) -- +(135.000000:8.485281pt and 8.485281pt) -- +(225.000000:8.485281pt and 8.485281pt) -- cycle;
\clip (68.000000, -0.000000) +(-45.000000:8.485281pt and 8.485281pt) -- +(45.000000:8.485281pt and 8.485281pt) -- +(135.000000:8.485281pt and 8.485281pt) -- +(225.000000:8.485281pt and 8.485281pt) -- cycle;
\draw (68.000000, -0.000000) node {{$\mathcal{F}^{\dagger}$}};
\end{scope}
\draw[color=black] (80.000000,15.000000) node[right] {$u_1$};
\draw[color=black] (80.000000,0.000000) node[right] {$f_2(u)$};
\end{tikzpicture}